# Integrating van der Waals materials on paper substrates for electrical and optical applications

*Wenliang Zhang[1], Qinghua Zhao[1,2,3], Carmen Munuera[1], Martin Lee[4], Eduardo Flores[5], João E. F. Rodrigues[1], Jose R. Ares[6], Carlos Sanchez[6,7], Javier Gainza[1], Herre S.J. van der Zant[4], José A. Alonso[1], Isabel J. Ferrer[6,7], Tao Wang[2,3], Riccardo Frisenda[1,\*], Andres Castellanos-Gomez[1,\*]*

[1] Materials Science Factory. Instituto de Ciencia de Materiales de Madrid (ICMM-CSIC), Madrid, E-28049, Spain.

[2] State Key Laboratory of Solidification Processing. Northwestern Polytechnical University. Xi'an, 710072, P. R. China

[3] Key Laboratory of Radiation Detection Materials and Devices. Ministry of Industry and Information Technology Xi'an, 710072, P. R. China

[4] Kavli Institute of Nanoscience, Delft University of Technology, Lorentzweg 1, 2628 CJ, Delft, The Netherlands.

[5] Centro de Nanociencias y Nanotecnología (CNyN), Universidad Nacional Autónoma de México (UNAM), km. 107, Carretera Tijuana-Ensenada, Ensenada, Baja California C.P. 22860, Mexico

[6] MIRE Group, Dpto. de Física de Materiales, Universidad Autónoma de Madrid, Madrid, E-28049, Spain

[7] Instituto Nicolás Cabrera, Universidad Autónoma de Madrid, UAM, Campus de Cantoblanco, E-28049 Madrid, Spain

**E-mail:** riccardo.frisenda@csic.es , andres.castellanos@csic.es

**ABSTRACT:**

Paper holds the promise to replace silicon substrates in applications like internet of things or disposable electronics that require ultra-low-cost electronic components and an environmentally friendly electronic waste management. In the last years, spurred by the abovementioned properties of paper as a substrate and the exceptional electronic, mechanical and optical properties of van der Waals (vdW) materials, many research groups have worked towards the integration of vdW materials-based devices on paper. Recently, a method to deposit a continuous film of densely packed interconnects of vdW materials on paper by simply rubbing the vdW crystals against the rough surface of paper has been presented. This method utilizes the weak interlayer vdW interactions and allows cleaving of the crystals into micro platelets through the abrasion against the paper. Here, we aim to illustrate the general character and the potential of this technique by fabricating films of 39 different vdW materials (including superconductors, semi-metals, semiconductors, and insulators) on standard copier paper. We have thoroughly characterized their optical properties showing their high optical quality: one can easily resolve the absorption band edge of semiconducting vdW materials and even the excitonic features present in some vdW materials with high exciton binding energy. We also measured the electrical resistivity for several vdW materials films on paper finding exceptionally low values, which are in some cases, orders of magnitude lower than those reported for analogous films produced by inkjet printing. We finally demonstrate the fabrication of field-effect devices with vdW materials on paper using the paper substrate as an ionic gate.

**MAIN TEXT:**

While the cost of silicon substrates ($\sim 1000$ € /m²) constitutes a big portion of the electronic components price tag, standard copier paper is incomparably inexpensive ($\sim 0.1$ € /m²).[1,2] Moreover, unlike silicon, paper is biodegradable and thus its use can relieve some of the urgent issues and challenges of the electronic waste management.[1,3–6] Indeed, although paper electronics cannot compete with silicon-based electronics in highly integrated circuits, simpler electronic components and sensors could be fabricated on paper substrates at a sizeable lower cost and with lower electronic waste footprint. Paper flexibility also makes it particularly suitable for flexible electronics applications.[5,7–10]

Given the fibrous structure of paper (that yields high surface roughness) the most widespread device fabrication approaches for van der Waals devices, based on mechanical exfoliation followed by lithographic and metal deposition steps, cannot be directly used. The scientific community, however, has developed new strategies to facilitate the fabrication of devices based on vdW materials onto paper substrates.[7,11,12] Up to now inkjet printing of inks prepared by liquid-phase exfoliation of vdW materials is probably the most common method and it has already demonstrated that it can be used to pattern complex devices, with high spatial resolution while combining different vdW materials.[13–16] Inkjet printing vdW materials on paper, however, is not exempt of open challenges.[17] First, this technique produces optimal results on the surface of specially engineered paper substrates, whose cost is substantially higher (5 - 15 € /m²) than standard copier paper,[5] and it needs non-standard printers. Another open challenge is that the drop-by-drop nature of inkjet printing makes it non-trivial to find the right parameters to obtain smooth and continuous films, avoiding the so-called "coffee ring" effect. Lastly, achieving highly conductive films using ink suspensions containing vdW materials is still challenging.[14,18–20]  This is attributed to the presence of insulating adsorbates, coming from the solvent of the ink suspension, on the surface of the deposited vdW materials thus preventing an intimate contact between platelets. These open challenges further motivated the development of alternative deposition methods to fabricate vdW materials on standard paper like the all-dry abrasion-induced deposition method. This method is based on the erosion of a vdW material while it is being rub against the rough surface of paper, similarly to the action of drawing/writing with pencil on paper. These abrasion forces break the weak vdW interlayer bonds, yielding a film formed by a network of interconnected platelets. This method has been used almost exclusively to fabricate sensors and devices based on graphite [21–29] but recently it has been extended to a handful of other vdW materials. [30–33] In this work we aim to demonstrate the general character of this technique, showing that it can be applied to fabricate dense films of many different vdW materials on standard paper. We have characterized the basic morphology, structural, optical and electrical properties of films of different vdW materials on paper to provide a basic playground for future works on paper-based electronics integrating dry-deposited vdW materials.

The deposition of the vdW materials is carried out by simply rubbing the desired material, in fine powder form, against the standard copier paper substrate with a cotton swab. Figure 1a shows a picture acquired during the deposition of a $MoSe_2$ film. Nitto tape (SPV 224) has been used to form a square shaped mask to control the geometry of the resulting film. Note that more complex mask designs can be realized by using a desktop vinyl cutter machine (Silhouette Portrait). Figure 1b shows an example of a paper substrate with a vinyl stencil mask, representing the logo of the '2D Foundry' research group, adhered on its surface. Different parts of the mask have been used to deposit different vdW materials: graphite, $MoS_2$, $WS_2$, $PbI_2$ and $AsSbS_3$. The resulting deposited vdW film, after peeling off the mask, is shown in the bottom panel of Figure 1b. Note that the desktop vinyl cutter machine allows for the patterning of features as small as $1\times1$ $mm^2$ and a minimum separation between features of 200 µm (see the Figure S1 in the Supporting Information).

We have found that this method to deposit vdW materials on paper is universal and it can be used to achieve densely packed films of a large variety of layered materials. Figure 1c shows a summary of 39 different vdW materials on paper. This catalogue contains diverse materials ranging from simple elemental materials (*e.g.* graphite, black phosphorus, tellurium…) to materials with very complex chemical and crystal structure (*e.g.* talc), and from superconductors (*e.g.* $NbSe_2$) to wide-gap insulators (*e.g.* h-BN, $MoO_3$). We address the reader to the Supporting Information (Figures S2 to S8) for a combined scanning electron microscopy (SEM) and energy-dispersive X-ray spectroscopy (EDX) analysis of the morphology and the chemical composition of the as-deposited films on paper. We observed that the vdW materials are homogeneously deposited over the paper surface except for few spots of uncoated paper that show up in the low magnification SEM images as brighter spots due to electrostatic charging. From the SEM images we also infer that abrasion-induced deposition process crushes the vdW materials flakes forming compact films onto the cellulose fibers while the gaps between fibers, where the pressure and friction forces are lower, are filled in with crystallites and it is sometimes possible to resolve loose flakes. Therefore, this deposition process leads to films with non-uniform thickness. This can be a handicap for applications, like high performance electronics or the fabrication of micro/nano devices, but it can be acceptable for other applications where the ultra-low cost of the substrate and its biodegradability are more important than the performance (*e.g.* smart tags and patches). The EDX spectra (Figures S2 to S8) also demonstrate that the abrasion-induced deposition does not modify the chemical composition of the vdW materials. Furthermore, Figures S9 and S10 of the Supporting Information show a Raman spectroscopy characterization of the as-deposited films on paper demonstrating that the layered materials are not structurally modified during the deposition process except for some laser-induced oxidation in few cases (e.g. $HfTe_2$). Characteristics phonon modes for the different films are resolved, and their identification performed based on previously reported data.

The optical properties of the as-prepared vdW films on paper were characterized by measuring their transmittance spectra in the 1.3 eV – 2.7 eV range (see Materials and Methods section for details about the measurement). Figure 2a shows a comparison of the spectra collected for different vdW materials on paper ranging from a semi-metal (graphite) to a wide-gap insulator (h-BN). The absorption band edge of the semiconducting materials with a gap value within the visible spectrum (*e.g.* $In_2Se_3$, $Sb_2S_3$, $As_2Se_3$, $AsSbS_3$, $As_2S_3$) is clearly visible as a rather abrupt increase in transmittance for energies lower than the band gap. The increase in transmittance matches very well with the band gap reported for these vdW materials in bulk and thin film.[34–37] Remarkably, some semiconducting vdW material films even show dips in the transmittance spectra associated with the generation of excitons. In fact, some transition metal dichalcogenides present high exciton binding energy that allow the observation of excitonic features in their reflectance and transmittance spectra even at room temperature.[38–43] Figure 2b illustrates how for $WS_2$, $MoS_2$, $WSe_2$, $MoSe_2$ and $ReS_2$ on paper one can still resolve those excitonic features (highlighted with an arrow), which are in good agreement with those observed for multilayer flakes.[38–43] These observations demonstrate the high optical quality of these vdW films on paper and opens the door to apply them in ultra-low-cost optical component as wavelength selective coating or filters. We address the Supporting Information (Figures S11 and S12) for the optical characterization of other 25 vdW materials on paper. Figure 2c shows an example where a $As_2S_3$ film on paper has been illuminated with monochromatic light with energy above and below its band gap showing a clear transition from opaque to transparent.

In order to study the electrical properties of the vdW material films we have used the transfer length method that consists of patterning the vdW material film under study into an elongated rectangular shape and depositing electrodes at increasing distance to extract the contact resistance and the resistivity. The electrodes are deposited by drawing with a graphite pencil (4B type, graphite content ~ 80%) rectangular contacts onto the vdW material film. Figure 3a shows a picture of a $WS_2$ film on paper with graphite electrodes separated from 1 mm to 7 mm. Note that, once the electrical measurements are accomplished, one can quantitatively measure the thickness of the film, which is needed to determine the resistivity, by slicing the device with a sharp razor blade and imaging the cross-section under an optical microscope (see Figure 3b). Figure 3c shows current *vs*. voltage characteristics (*IV*s hereafter) measured between different pair of electrodes with spacing ranging from 1 mm to 7 mm on a $WS_2$ film. The *IV*s are very linear in the ±1 V voltage range and even for much higher voltages (see Figure S13 in the Supporting Information for *IV*s in ±30 V range). Figure 3d shows the resistance values (extracted from the slopes of the *IV*s in Figure 3c) as a function of the electrode distance that follows a linear trend. The intercept of that line with the vertical axis gives twice the contact resistance ($R_c$) value and the slope of the trend ($R/L$) allows determining the resistivity ($\rho$) as it is defined as:

$$\rho = R \cdot W \cdot th \, / \, L$$

where $R$ is the resistance and $L$, $W$ and $th$ are the length, width and thickness of the channel respectively.

From the transfer length measurement for $WS_2$ on paper, shown in Figure 3d, we obtain a contact resistance $R_c \sim 150$ k$\Omega$ and a resistivity $\rho$ = 440 ± 10 $\Omega \cdot$m. We have carried out similar transfer length measurements on other 4 $WS_2$ devices, finding resistivity values between $\sim 360$ $\Omega \cdot$m and $\sim 530$ $\Omega \cdot$m (see the Supporting Information, Figure S16). These resistivity values are remarkably low as compared to the best resistivity values reported in the literature for films prepared with inks of liquid phase exfoliated $WS_2$ ($10^4$ $\Omega \cdot$m).[44] Note that we have also measured six other transfer length devices with $WS_2$ from other two material sources, finding resistivity values in the $\sim 70 - 370$ $\Omega \cdot$m range. See the Supporting Information (Figure S16) for the datasets corresponding to the other two $WS_2$ sources.

We attribute these low resistivity values to the all-dry nature of the deposition method which avoids the presence of inter-platelet adsorbates that can be difficult to remove and may hamper the electrical transport. To get an insight about the device reproducibility we have tested 118 $WS_2$ devices with 2 mm of channel length and 10 mm of width. Figure 3e shows a histogram with the resistance values measured on the 118 $WS_2$ devices. The histogram follows a skewed normal distribution with a peak at $\sim 3$ M$\Omega$. Interestingly $\sim 95\%$ of the devices have resistances in the $1.0 - 6.7$ M$\Omega$, a low dispersion taking into account the nature of the films: a random network of interconnected platelets where percolation transport is expected. In fact, we have simulated the percolative electronic transport of the van der Waals devices on paper through a random resistor network model finding very similar statistical (a lognormal) distribution of the resistance (see the Figure S19 in the Supporting Information for details about the simulation). See Figure S20 in the Supporting Information for a calculation of the resistance dispersion as a function of channel length in a percolative film. The inset in Figure 3e shows a histogram of the average thickness values measured on 35 $WS_2$ films finding that the thickness of the deposited vdW films have a median value of $\sim 20 \pm 5$ μm. (see the Supporting Information Figure S21 for a thickness histogram including the thickness of other vdW materials films).

Figure 3f summarizes the resistivity values measured in transfer length devices on paper based on 15 different vdW materials: $NbSe_2$, graphite (from 4 different sources), GeTe, $TiS_3$, $TiS_2$, SnSe, SnS, BP, $WS_2$ (from 3 different sources), $MoSe_2$, $WSe_2$, $MoS_2$ (from 3 different sources), $ReS_2$, $In_2Se_3$ and GaTe. These materials range from metals/semimetals to semiconductors. We obtain resistivity values as low as $(2.1 - 5.6) \cdot 10^{-3}$ Ω·m for $NbSe_2$, $1.7 \cdot 10^{-3} - 4.7 \cdot 10^{-2}$ Ω·m for graphite (depending on the graphite source) and $8.2 \cdot 10^{-2} - 1.2 \cdot 10^{-1}$ Ω·m for GeTe, metal or semimetal vdW materials. For narrow gap semiconducting materials ($TiS_3$, $TiS_2$, SnSe, SnS and BP) we obtain values spanning from 4.1 Ω·m for $TiS_3$ to $3.0 \cdot 10^4$ Ω·m (for BP). For the semiconducting materials with a band gap in the visible part of the spectrum ($MoSe_2$, $WSe_2$, $MoS_2$, $ReS_2$, $In_2Se_3$ and GaTe) all the resistivity values obtained are between $2.5 \cdot 10^4$ and $3.9 \cdot 10^5$ Ω·m, in the same range of the lowest resistivity values reported for inkjet-printed $MoS_2$ films ($\sim 1.1 \cdot 10^4 - 1.7 \cdot 10^6$ Ω·m).[14,18–20] For $WS_2$, as discussed above, we obtain lower resistivity values (in the 72.6 – 528.2 Ω·m). These measurements further demonstrate how the all-dry nature of the abrasion-based deposition method allows the fabrication of films with high electrical performance. We address the reader to the Supporting Information for the datasets of the transfer length devices, measured for the different vdW materials (Figures S14, S15, S16, S17 and S18). The $WS_2$ resistance histogram, shown in Figure 3e, could be used as an estimation for the standard variation expected for the resistivity of these films: $\sim \pm 33$-50%. Note that one would expect that the source of powdered vdW material could have a strong impact in the properties of the deposited films. This is still an open problem in the community working on van der Waals materials. There are only few works devoted to systematically correlate the properties of exfoliated materials produced from different sources of van der Waals materials.[45–47] Here we compared the resistivity values obtained for graphite, $WS_2$ and $MoS_2$ from different sources finding that sources with smaller particle size led to lower electrical resistivity. But a more comprehensive study in a single vdW material should be done in the future to draw more robust conclusions.

We further explore the possibility of fabricating more complicated devices such as field-effect devices with the films of vdW materials on paper. To do so we employ a back-gate configuration (see Figure 4a)[48] where the fillers (e.g. $CaCO_3$) and hydroxyl groups lead to the formation of cation-anion pairs upon electrical field bias.[49–51] Therefore, the paper substrate acts as a solid ionic gate that can be used to modulate the electrical properties of the as deposited vdW materials. Figure 4b shows the conductance of 4 paper-supported field-effect devices based on different vdW materials (graphite, $TiS_2$, $WS_2$ and $TiS_3$) as a function of the voltage applied to the gate electrode. The insets show the current *vs*. voltage characteristics acquired at different gate voltages. While graphite shows a weak ambipolar (although stronger for hole) modulation of the conductance, the other materials show a sizeable change in the conductance upon gating. This is expected due to the semi-metallic nature of graphite which leads to a more effective electric field screening than for semiconducting materials. $TiS_2$ and $TiS_3$ are n-type and show modulations up to a factor of ~8. $WS_2$, on the other hand, shows a p-type behaviour with a modulation of the conductance up to a factor of ~19. These results illustrate the feasibility of fabricating more complex electronic devices on paper with vdW materials.

**Conclusions**

In summary, we demonstrate the potential of abrasion-induced deposition method to fabricate films of a large variety of vdW materials, with different structural, optical and electrical properties, on standard copier paper. We characterize the optical properties of 39 vdW material films by measuring their transmittance spectra, finding that the semiconducting films show well-defined absorption band edges and, in some cases, even excitonic features. The electrical properties of $WS_2$ films have been thoroughly characterized by studying more than 100 devices finding resistivities in the 70 – 530 $\Omega \cdot m$ range, which is about 20 –150 times lower than the lowest reported resistivity values for films based on $WS_2$ inks prepared by liquid phase exfoliation. We have also measured the resistivities of 14 additional vdW materials with different properties ranging from metal/semimetals to semiconductors with a sizeable gap. A resistivity as low as $2.1 \cdot 10^{-3}$ $\Omega \cdot m$ has been measured for metallic $NbSe_2$. We also demonstrate the possibility of fabricating field-effect devices with vdW materials on paper, with a gate dependent modulation of the conductance up to a factor of $\sim$19, using the paper substrate as an ionic gate. Our results show how the abrasion-induced deposition method yields vdW material films on paper capable of high performance for optical and electrical applications and open the door to further work exploring the intrinsic properties and applications of vdW materials on standard paper substrates.

**Materials and Methods**

*Materials sources.*

Graphite was extracted from 4B and 8B Faber Castell pencils. Nano graphite powder with 400 nm (PN: MKN-CG-400) and 50 nm (PN: MKN-CG-50) average particle sizes were purchased at Lowerfriction Lubricants.

Antimony (Smart-elements), bismuth (Novaelements) and tellurium (Novaelements) high purity (>99.99%) chunks were crushed and manually ground before abrasion-induced deposition on paper.

BP (black phosphorus) was synthesized using a high-pressure procedure in a piston-cylinder press (Rockland Research Co.), and a pressure of 2 GPa at high temperature of 1073 K for 1 h. Initially, small pieces of amorphous red phosphorus (Alfa-Aesar) were ground, in an agate mortar inside a nitrogen-filled glove box; the material was sealed in a niobium capsule, and then introduced inside a cylindrical graphite heater. After quenching and releasing pressure, pieces of BP (6–7 mm in diameter; 5 mm in thickness) were recovered.[52] The as-synthesized BP chunks were crushed and manually ground before abrasion-induced deposition on paper.

hBN (PN: 11078.18), $SnS$ (PN: 14051.06), $SnSe$ (PN: 18781.06), $GeTe$ (PN: 45461.03), $WSe_2$ (PN: 13084.09), $TiS_2$ (PN: 12826.06), $NbSe_2$ (PN: 13101.09), $MoSe_2$ (PN: 13112.06), $HfTe_2$ (PN: 39223.03), $ReS_2$ (PN: 89482.04), $Sb_2Se_3$ (PN: 13130.06), $Sb_2Te_3$ (PN: 45922.09), $In_2Se_3$ (PN: 88280.06), $As_2Se_3$ (PN: 13130.06), $Bi_2Se_3$ (PN: 47198.06), $Bi_2Te_3$ (PN: 44077.06), $MoO_3$ (PN: A11159.18) high purity materials were purchased from Alfa Aesar. Most materials came in fine powder form, but some materials had to be manually ground before using.

High purity $PbI_2$ was purchased at Sigma Aldrich (PN: 203602-50G) and re-crystallized into micro-platelets shape.[53]

BPSCCO ($Bi_{1.6}Pb_{0.4}Sr_{1.6}Ca_{2.0}Cu_{2.8}O_{9.2+x}$) fine powder was purchased from Sigma Aldrich (PN: 378720-10G).

$MoS_2$ of three different sources were used in this work: Alfa Aesar (PN: 41827.14), Hagen Automation Ltd and a natural molybdenite mineral.

$WS_2$ of three different suppliers were used in this work: Hagen Automation Ltd, Alroko GmbH & Co KG and Alfa Aesar (PN: 11829.18)

InSe, GaSe, GaTe single crystals were grown by the Bridgman method[54–57]. High purity In (7N), Ga (6N), Se (6N) and Te (7N) was used as the raw material. Gallium was baked at 673K for 4h under high vacuum to remove the oxide layer. Synthesis were performed before crystal growth. For GaTe, direct synthesis was used in a rocking furnace where Ga and Te were mixed in a stoichiometric ratio and sealed in a quartz crucible at $10^{-5}$ Pa. For InSe and GaSe, Physical Vapor Transfer (PVT) method was used. In/Ga and Se source were placed at both ends of a horizontal crucible in a nine-zone furnace. The source temperature was optimized separately during the synthesis and cooling process to ensure the stoichiometry of the compound. After that, a six-temperature zone furnace was used to grow the InSe/GaSe/GaTe single crystal by vertical Bridgman method. The temperature gradient is usually 5-10 K/cm, the growth rate is among 0.5-1 mm/h. T

$Sb_2S_3$, $As_2S_3$, $AsSbS_3$, $PbSnS_2$, franckeite, cylindrite and talc materials were extracted from natural minerals directly purchased in mineral collector's shops. The minerals were manually ground before using.

$TiS_3$, $ZrS_3$, $HfS_3$: Powders of $TiS_3$, $ZrS_3$ and $HfS_3$ were synthesized by a solid–gas reaction of metal powders of Ti (Goodfellow, 99.5%), Zr (Johnson Matthey, 99.9%) and Hf (Alfa Aesar, 99.6%) with sulfur (Merck, 99.75%) at molar ratios of M/S =3 in a vacuum sealed ampoule annealed at 550 °C during 60 h. To obtain $TiS_2$ powder the annealing temperature used was 600 °C and the Ti/S molar ratio = 2.[58]

*Scanning electron microscopy and energy-dispersive X ray spectroscopy characterization.* The topography and the chemical composition of the films of vdW materials on paper were characterized using a FEI Helios G4 CX system. A thin film of sputtered gold or a spin coated layer of electrically conductive resist (Electra 92) were used to image films of insulating vdW materials (e.g. hBN, $MoO_3$ or talc).

*Raman characterization.* Raman spectra of the vdW materials were acquired under ambient conditions using a confocal Raman microscope (MonoVista CRS+, Spectroscopy & Imaging GmbH, Germany) with a 1500 lines/mm grating in the backscattering geometry. A 532 nm line of a CW solid-state laser at 0.5 mW power was used for excitation through a 100× magnification microscope objective (NA = 0.9). The 300 lines/mm grating was used to acquire a reference spectrum of the paper substrate, covering the full Raman spectrum (50 to 4600cm-1).

*Optical characterization.* The transmittance spectroscopy measurements were carried out by illuminating the samples with fiber coupled halogen lamp (OSL2, Thorlabs) and collecting the transmitted light with a fiber coupled CCD spectrometer (CCS200/M, Thorlabs). The transmittance spectra ($T$) were obtained by making the quotient of the transmission spectrum acquired on the blank paper ($I_0$) and that obtained on the paper covered by the vdW film ($I$): $T = I/I_0$.

*Electronic characterization.* The electrical transport measurements were carried out in a homebuilt probe station operated at atmospheric conditions. A Keithley 2450 source-measure unit was used to acquire the current *vs.* voltage characteristics. For the field effect measurements, a pair of programmable benchtop power supply (TENMA 72-2545) were connected to the back-gate electrode.

**ACKNOWLEDGEMENTS**

We thank Prof. Jonathan Coleman (Trinity College Dublin) for fruitful discussions about the electrical properties of networks based on interconnected 2D materials. This project has received funding from the European Research Council (ERC) under the European Union's Horizon 2020 research and innovation program (grant agreement n° 755655, ERC-StG 2017 project 2D-TOPSENSE) and the European Union's Horizon 2020 research and innovation program under the Graphene Flagship (grant agreement number 785219, GrapheneCore2 project and grant agreement number 881603, GrapheneCore3 project).

R.F. acknowledges the support from the Spanish Ministry of Economy, Industry and Competitiveness (MINECO) through a Juan de la Cierva-formación fellowship 2017 FJCI-2017-32919. J.G. and J.A.A thank the MINECO for finding the project MAT2017-84496-R. We also acknowledge funding from the Spanish Ministry of Science, Innovation and Universities: RTI2018-099794-B-100. W. Zhang acknowledges the grant from China Scholarship Council (CSC) under No. 201908610178. Q.H.Z. acknowledges the grant from China Scholarship Council (CSC) under No. 201706290035.

**COMPETING INTERESTS**

The authors declare no competing financial interests.

**FUNDING**

European Research Council (ERC): grant agreement n° 755655

European Union's Horizon 2020 research and innovation program: grant agreement number 785219, grant agreement number 881603

Spanish Ministry of Economy, Industry and Competitiveness: 2017 FJCI-2017-32919.

Spanish Ministry of Science, Innovation and Universities: RTI2018-099794-B-100

China Scholarship Council: 201908610178 and 201706290035

**DATA AVAILABILITY:**

The raw/processed data required to reproduce these findings cannot be shared at this time due to technical or time limitations.

Data will be made available on request.

**FIGURES**

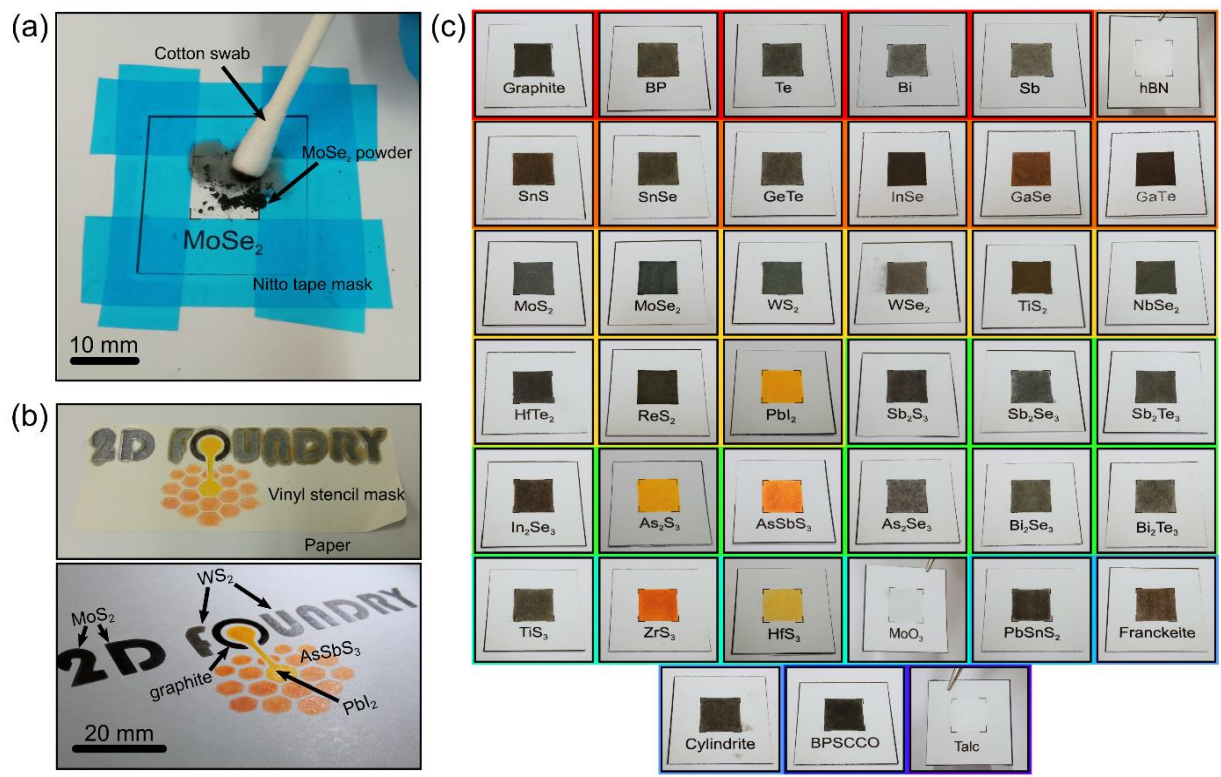

**Figure 1. Deposition of vdW materials on paper by the abrasion-based method.** (a) Picture of the process employed to deposit a $MoSe_2$ film on standard paper by mechanically rubbing $MoSe_2$ powder against the surface of paper with a cotton swab. (b) Example of the use of a custom-designed vinyl stencil mask to deposit the vdW materials on paper following a desired geometry and even allowing the deposition of different vdW materials on different areas of the paper. (c) Pictures of the catalogue of vdW materials deposited on paper by the abrasion-based method to illustrate the general character of this technique.

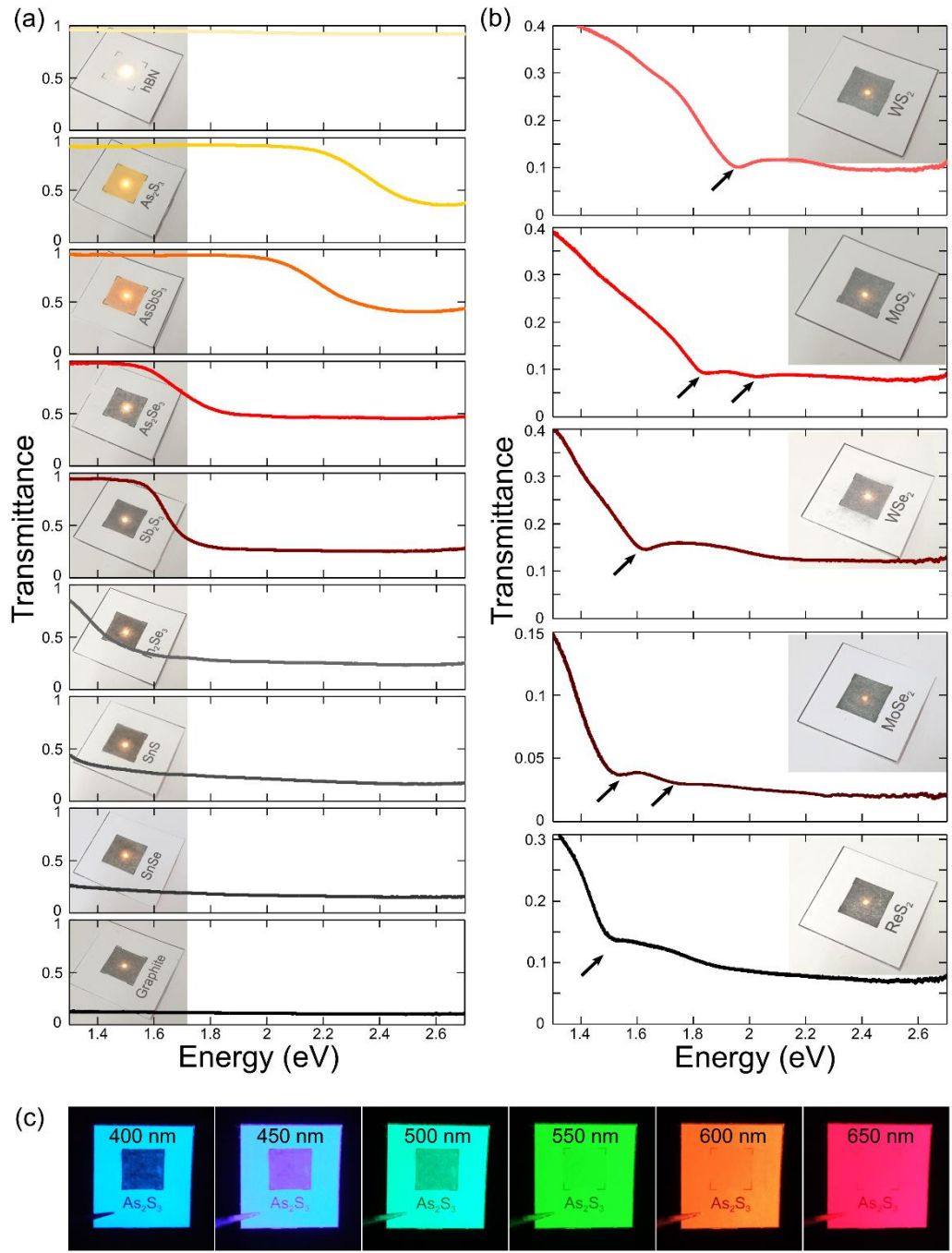

**Figure 2. Optical properties of vdW materials on paper.** (a) Transmittance spectra of 9 different vdW materials films on paper with band gaps ranging from 0 eV (graphite) to ~6 eV (hBN). For $In_2Se_3$, $Sb_2S_3$, $As_2Se_3$, $AsSbS_3$ and $As_2S_3$ the absorption band edge can be resolved in the measured spectra as an abrupt change of the transmittance. (b) Transmittance spectra of 5 vdW materials on paper with a high exciton binding energy that present dips (highlighted with arrows) in the spectra associated to the photogeneration of excitons. The energy of the dips matches with the literature values of the excitons obtained on multilayered flakes through photoluminescence, reflectance and absorption spectroscopies. (c) Pictures of a $As_2S_3$ film illuminated with monochromatic light with energy higher than the band gap ($\lambda$ < 520 nm) and lower than the band gap ($\lambda$ > 520 nm) where a sudden change in opacity of the film is observed.

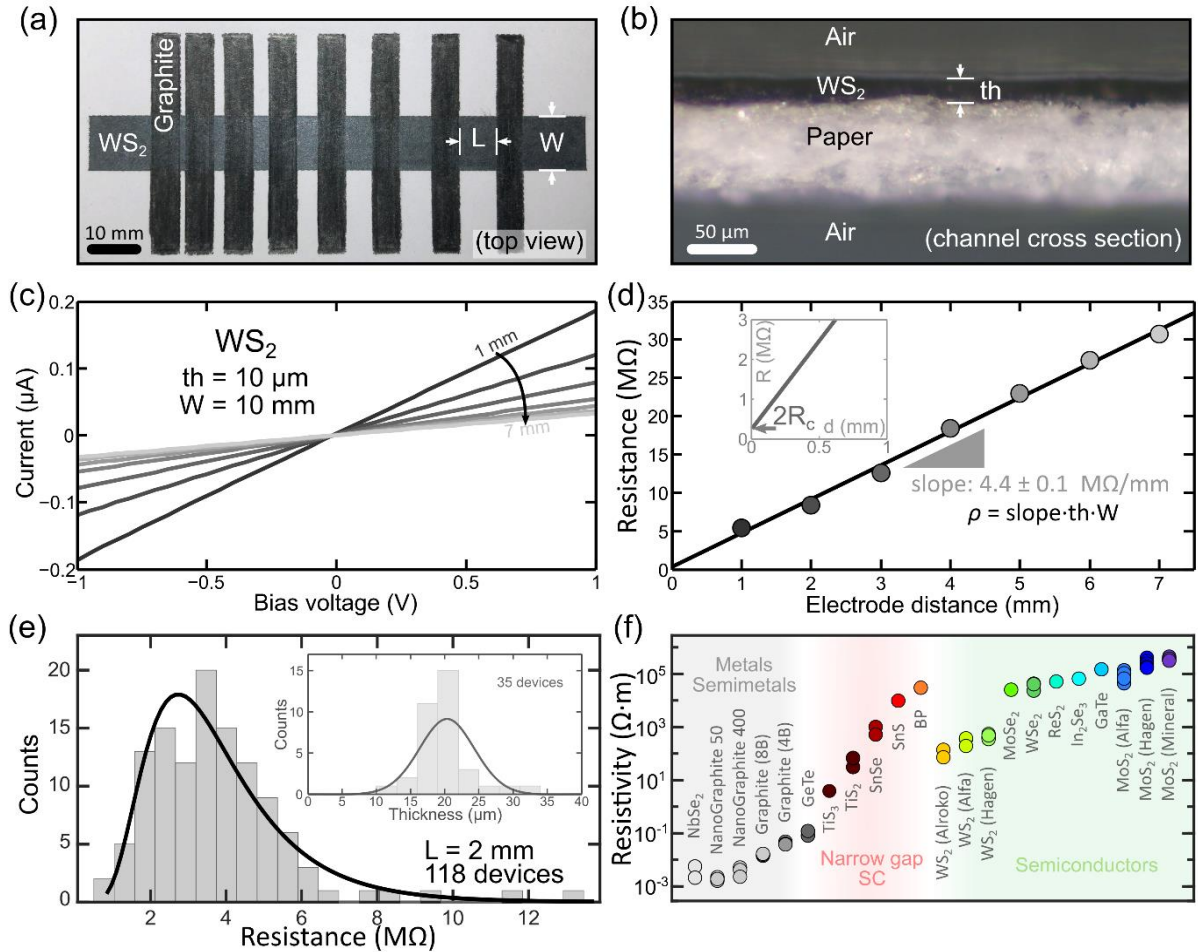

**Figure 3. Electrical properties of vdW materials on paper.** (a) Picture of a transfer length device on standard paper with a $WS_2$ channel and graphite electrodes. (b) Optical microscopy image of the cross-section of the $WS_2$ film on paper where the average thickness of the $WS_2$ film can be measured. (c) Current vs. voltage characteristics (*IVs*) of $WS_2$ channels with different length from 1 mm to 7 mm. (d) Resistance vs. channel length, extracted from the *IVs* in (c). The slope of the resistance vs. length dataset can be used to extract the resistivity and the intercept with the vertical axis allows the extraction of the contact resistance $R_c$ (see inset). (e) Histogram constructed with the resistance values measured on 118 $WS_2$ devices ($L$ = 2 mm, $W$ = 10 mm) to get an estimation of the device-to-device variation. A well-defined peak at ~2-4 MΩ is obtained. The inset in (e) shows a histogram of the thickness values measured on 35 $WS_2$ films on paper where a well-defined average thickness of ~20 ± 5 µm is obtained. (f) Comparison of the resistivity values measured for films of 15 different vdW materials.

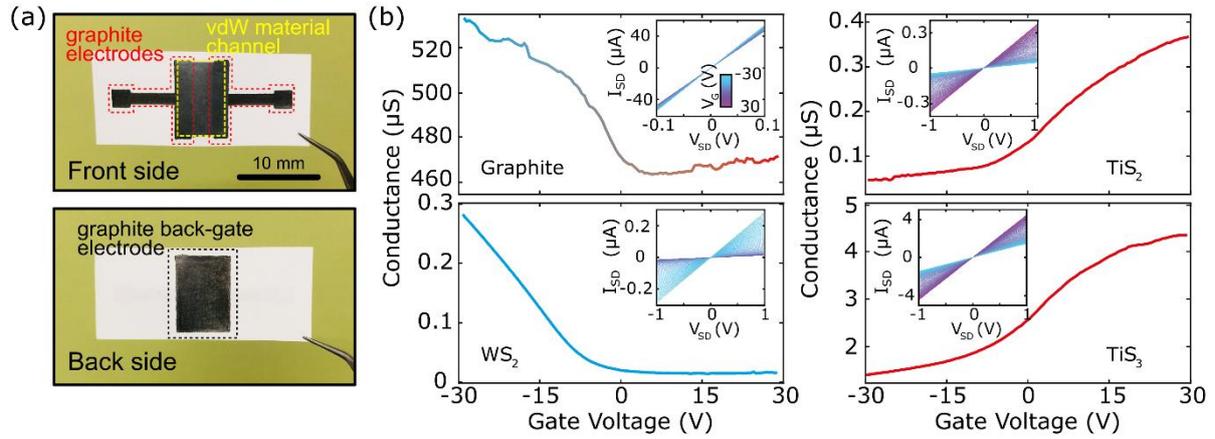

**Figure 4. Electric field-effect in vdW materials on paper.** (a) Pictures of a device employed in electric field-effect gating of van der Waals films on paper. A graphite back-gate electrode is deposited on the back side of the paper substrate. (b) Conductance of graphite (Nanographite 400), $TiS_2$, $TiS_3$ and $WS_2$ (Hagen Automation) as a function of the voltage applied to the back-gate electrode. The insets show the current *vs*. voltage characteristics of the devices at different gates.

**Supporting Information:**

# Integrating van der Waals materials on paper substrates for electrical and optical applications

*Wenliang Zhang[1], Qinghua Zhao[1,2,3], Carmen Munuera[1], Martin Lee[4], Eduardo Flores[5], João E. F. Rodrigues[1], Jose R. Ares[6], Carlos Sanchez[6,7], Javier Gainza[1], Herre S.J. van der Zant[4], José A. Alonso[1], Isabel J. Ferrer[6,7], Tao Wang[2,3], Riccardo Frisenda[1,\*], Andres Castellanos-Gomez[1,\*]*

[1] Materials Science Factory. Instituto de Ciencia de Materiales de Madrid (ICMM-CSIC), Madrid, E-28049, Spain.

[2] State Key Laboratory of Solidification Processing. Northwestern Polytechnical University. Xi'an, 710072, P. R. China

[3] Key Laboratory of Radiation Detection Materials and Devices. Ministry of Industry and Information Technology Xi'an, 710072, P. R. China

[4] Kavli Institute of Nanoscience, Delft University of Technology, Lorentzweg 1, 2628 CJ, Delft, The Netherlands.

[5] Centro de Nanociencias y Nanotecnología (CNyN), Universidad Nacional Autónoma de México (UNAM), km. 107, Carretera Tijuana-Ensenada, Ensenada, Baja California C.P. 22860, Mexico

[6] MIRE Group, Dpto. de Física de Materiales, Universidad Autónoma de Madrid, Madrid, E- 28049, Spain

[7] Instituto Nicolás Cabrera, Universidad Autónoma de Madrid, UAM, Campus de Cantoblanco, E-28049 Madrid, Spain

**Resolution of deposition through vinyl stencil masks**

**Scanning electron microscopy and energy-dispersive X-ray spectroscopy characterization of vdW materials on paper**

**Raman characterization of vdW materials on paper**

**Transmittance spectra of the vdW materials not shown in the main text**

**Current *vs*. voltage characteristics for high voltage biasing**

**Electrical characterization of 14 vdW materials on paper**

**Comparison of WS$_2$ devices from different material sources**

**Comparison of MoS$_2$ devices from different material sources**

**Comparison of graphite devices from different material sources**

**Simulation of a network of interconnected platelets through a random resistor network model**

**Histogram of the thickness of different vdW materials films on paper**

**Resolution of deposition through vinyl stencil masks:**

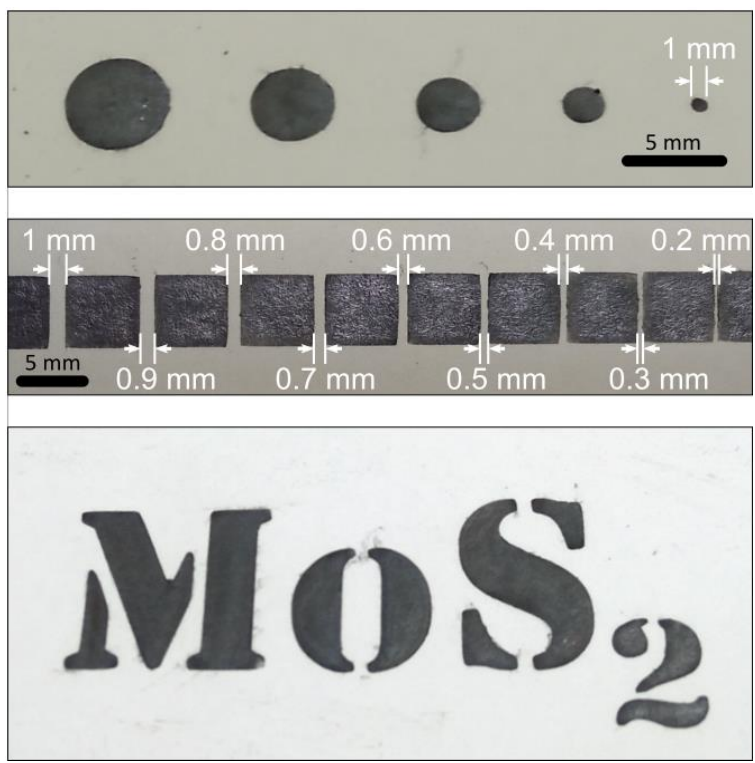

**Figure S1. Resolution of the custom-design vinyl stencil mask based deposition.** (top) Picture of MoS$_2$ circles deposited on standard paper through a vinyl stencil mask where a minimum feature of 1 mm is achieved. (middle) Picture of 5×5 mm$^2$ MoS$_2$ squares deposited on paper through a vinyl stencil mask with different spacing between neighboring squares. A minimum inter-square distance of 200 μm is achieved. (Bottom) Picture of "MoS$_2$" text 'drawn' with MoS$_2$ on paper using a vinyl stencil mask.

**Scanning electron microscopy and energy-dispersive X-ray spectroscopy characterization of vdW materials on paper:**

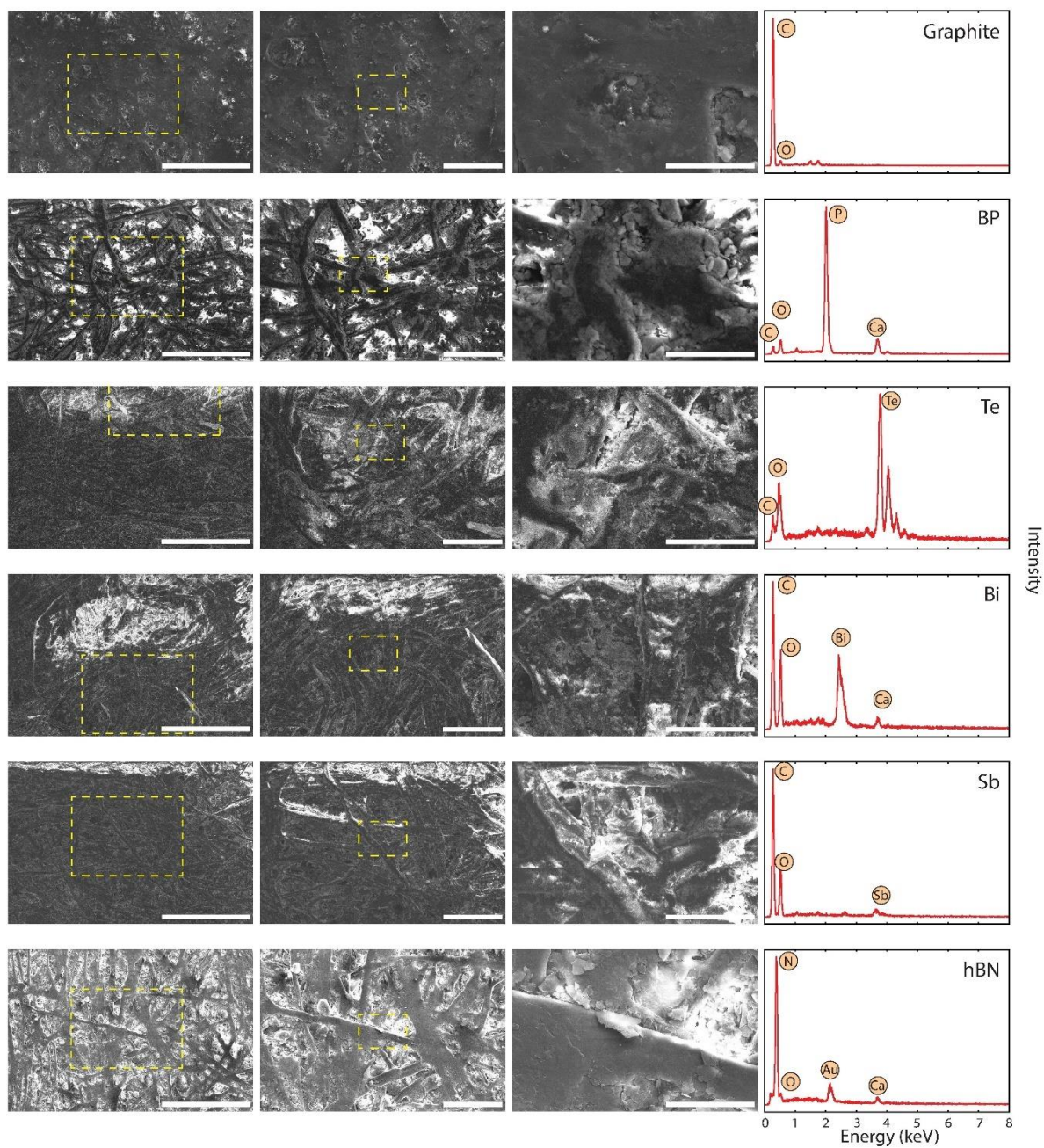

**Figure S2. SEM and EDX characterization of different films of vdW materials on paper.** Each row included 3 SEM images at increasing magnification and an EDX spectra of a different vdW material on paper (labelled in the EDX spectra panel). A thin layer of sputtered gold has been used in some highly resistive samples to avoid electrostatic charging effects; this might show up in the EDX spectra as a 'Au' peak. Some EDX spectra also show peaks associated to calcium presence. These peaks are due to the underlying paper and arise from the calcium carbonate (a bright white mineral with chemical formula $CaCO_3$) added as a filler in the paper pulp. Scale bars are 300 μm, 100 μm and 30 μm from leftmost panel to the rightmost panel.

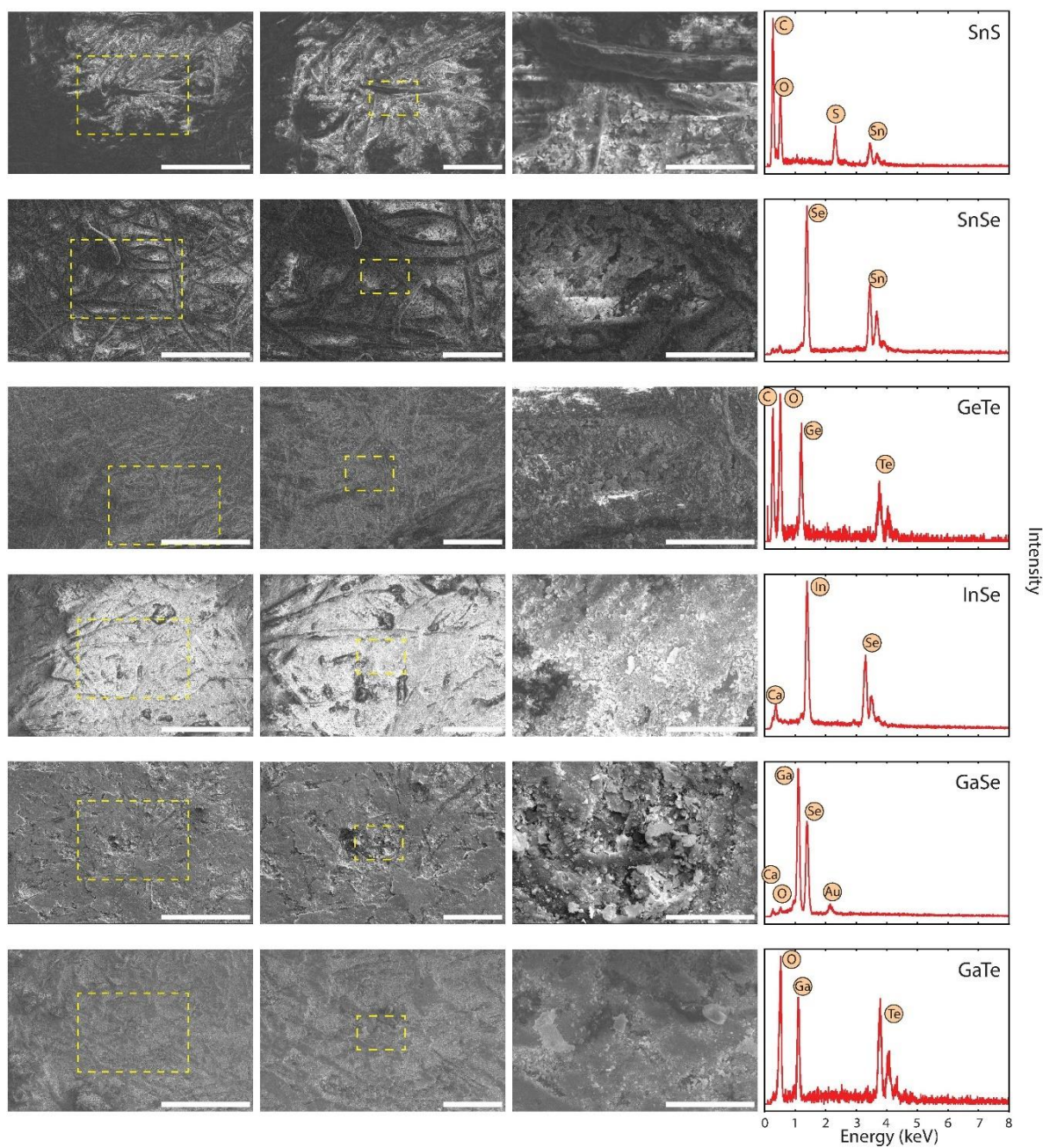

**Figure S3. SEM and EDX characterization of different films of vdW materials on paper.** Each row included 3 SEM images at increasing magnification and an EDX spectra of a different vdW material on paper (labelled in the EDX spectra panel). A thin layer of sputtered gold has been used in some highly resistive samples to avoid electrostatic charging effects; this might show up in the EDX spectra as a 'Au' peak. Some EDX spectra also show peaks associated to calcium presence. These peaks are due to the underlying paper and arise from the calcium carbonate (a bright white mineral with chemical formula $CaCO_3$) added as a filler in the paper pulp. Scale bars are 300 μm, 100 μm and 30 μm from leftmost panel to the rightmost panel.

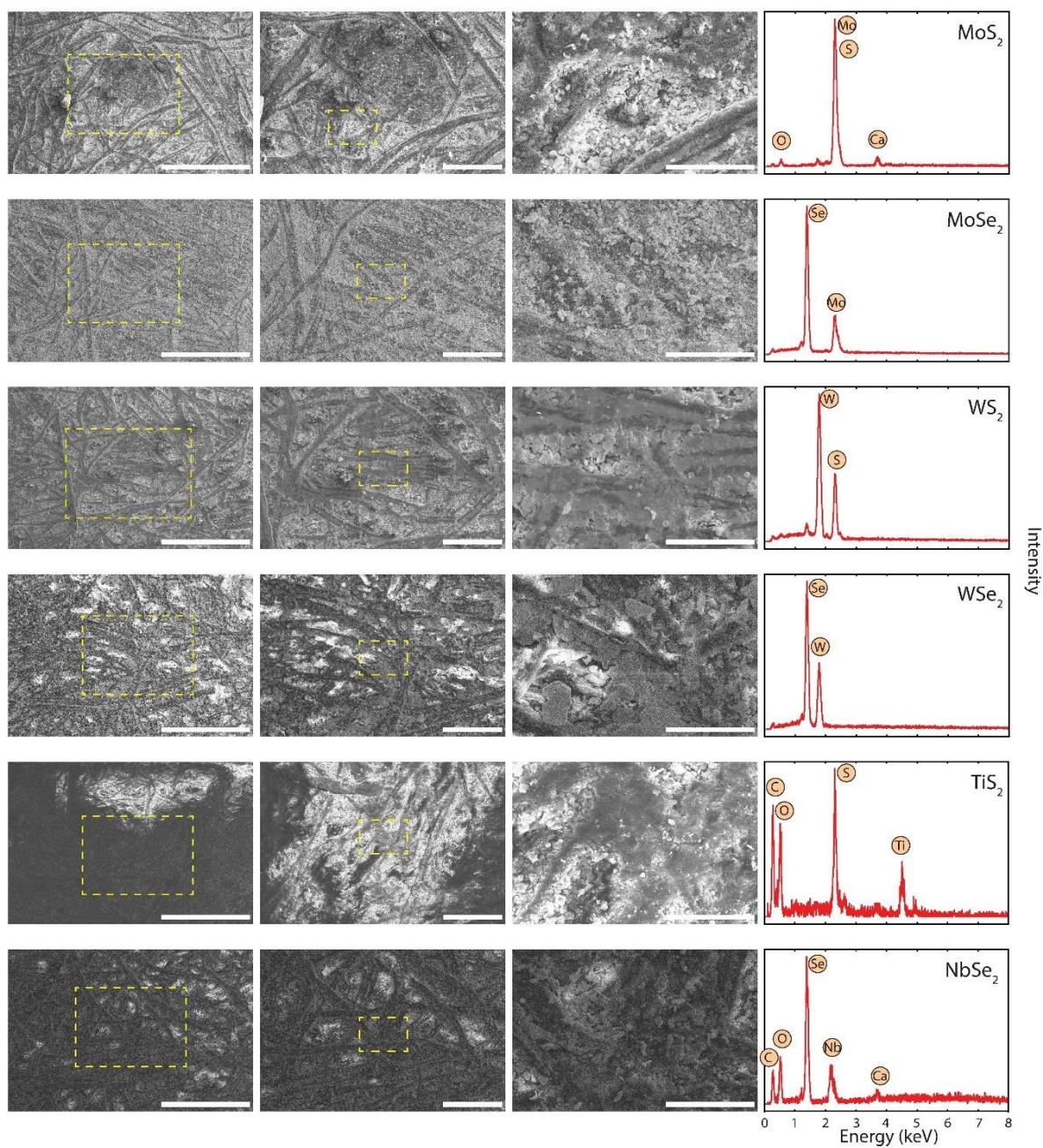

**Figure S4. SEM and EDX characterization of different films of vdW materials on paper.** Each row included 3 SEM images at increasing magnification and an EDX spectra of a different vdW material on paper (labelled in the EDX spectra panel). A thin layer of sputtered gold has been used in some highly resistive samples to avoid electrostatic charging effects; this might show up in the EDX spectra as a 'Au' peak. Some EDX spectra also show peaks associated to calcium presence. These peaks are due to the underlying paper and arise from the calcium carbonate (a bright white mineral with chemical formula $CaCO_3$) added as a filler in the paper pulp. Scale bars are 300 μm, 100 μm and 30 μm from leftmost panel to the rightmost panel.

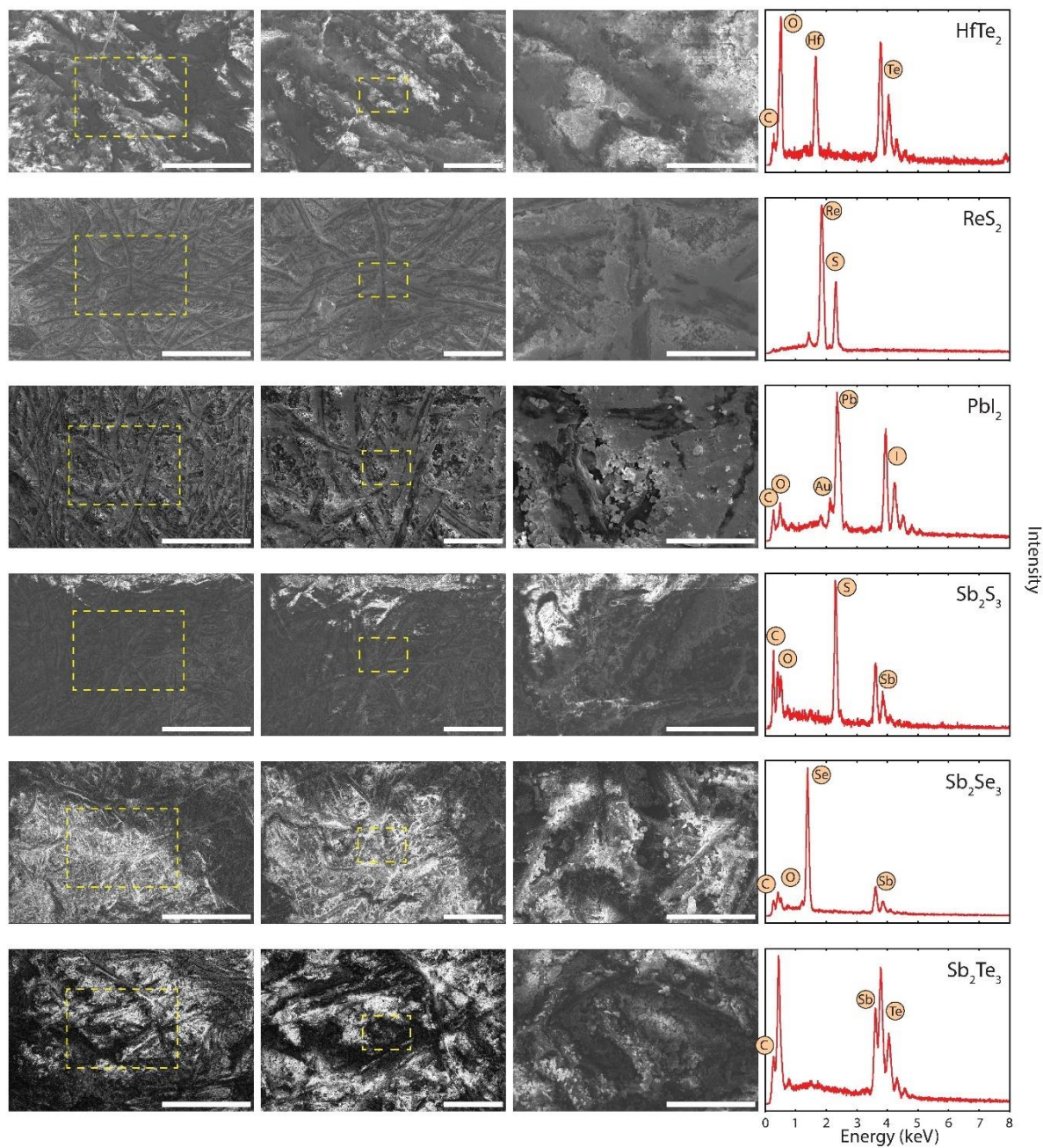

**Figure S5. SEM and EDX characterization of different films of vdW materials on paper.** Each row included 3 SEM images at increasing magnification and an EDX spectra of a different vdW material on paper (labelled in the EDX spectra panel). A thin layer of sputtered gold has been used in some highly resistive samples to avoid electrostatic charging effects; this might show up in the EDX spectra as a 'Au' peak. Some EDX spectra also show peaks associated to calcium presence. These peaks are due to the underlying paper and arise from the calcium carbonate (a bright white mineral with chemical formula $CaCO_3$) added as a filler in the paper pulp. Scale bars are 300 μm, 100 μm and 30 μm from leftmost panel to the rightmost panel.

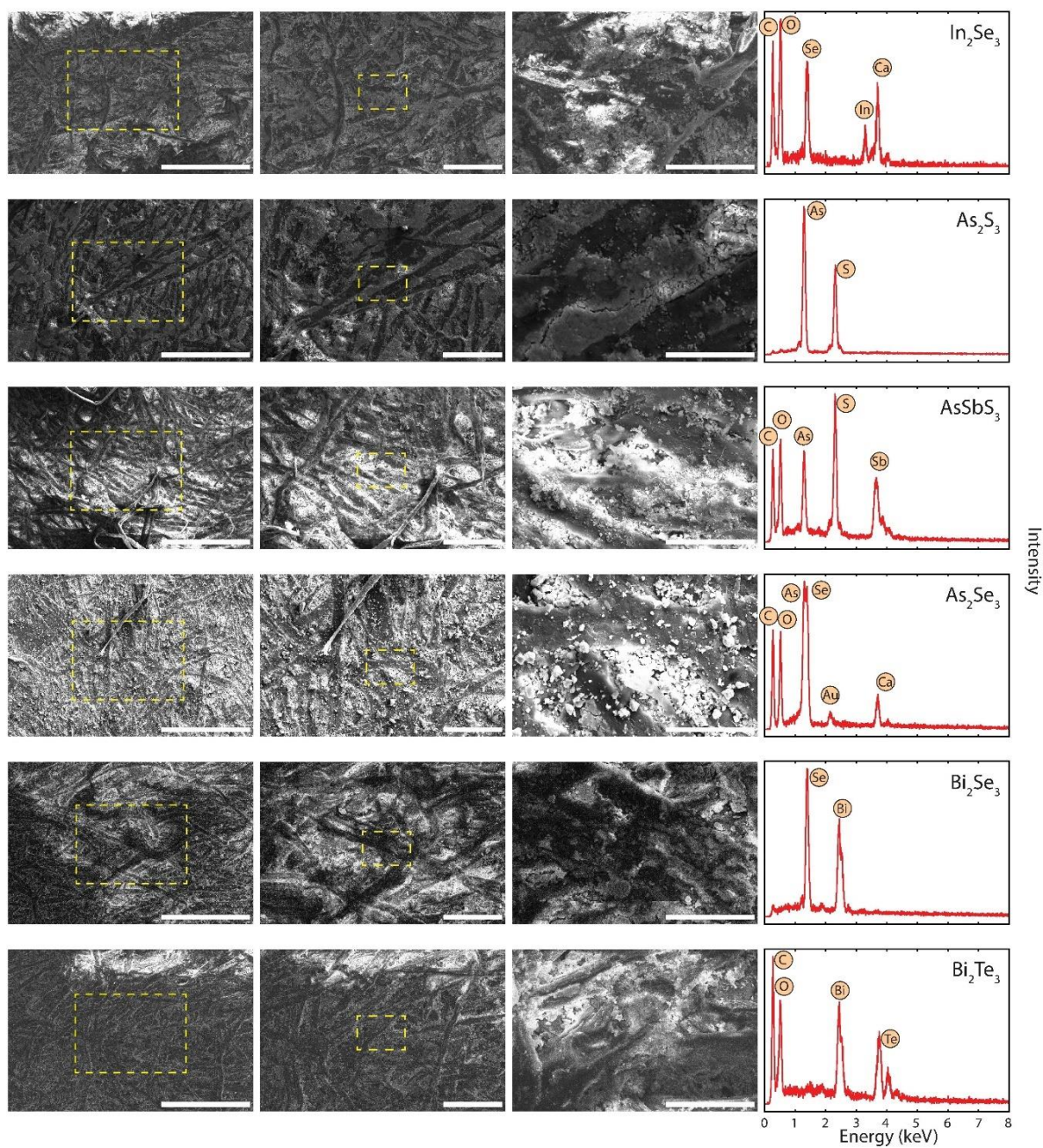

**Figure S6. SEM and EDX characterization of different films of vdW materials on paper.** Each row included 3 SEM images at increasing magnification and an EDX spectra of a different vdW material on paper (labelled in the EDX spectra panel). A thin layer of sputtered gold has been used in some highly resistive samples to avoid electrostatic charging effects; this might show up in the EDX spectra as a 'Au' peak. Some EDX spectra also show peaks associated to calcium presence. These peaks are due to the underlying paper and arise from the calcium carbonate (a bright white mineral with chemical formula $CaCO_3$) added as a filler in the paper pulp. Scale bars are 300 μm, 100 μm and 30 μm from leftmost panel to the rightmost panel.

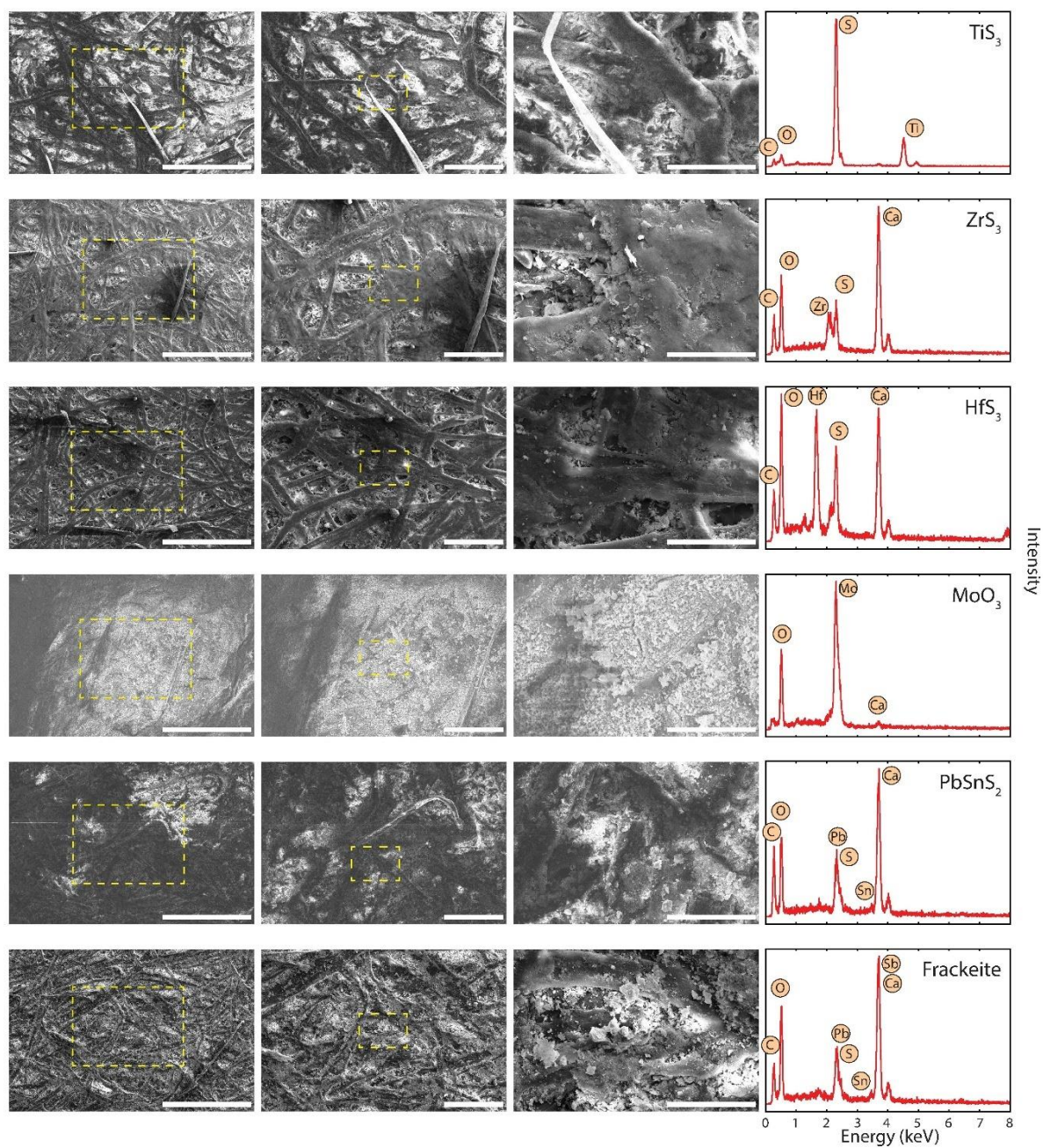

**Figure S7. SEM and EDX characterization of different films of vdW materials on paper.** Each row included 3 SEM images at increasing magnification and an EDX spectra of a different vdW material on paper (labelled in the EDX spectra panel). A thin layer of sputtered gold has been used in some highly resistive samples to avoid electrostatic charging effects; this might show up in the EDX spectra as a 'Au' peak. Some EDX spectra also show peaks associated to calcium presence. These peaks are due to the underlying paper and arise from the calcium carbonate (a bright white mineral with chemical formula $CaCO_3$) added as a filler in the paper pulp. Scale bars are 300 µm, 100 µm and 30 µm from leftmost panel to the rightmost panel.

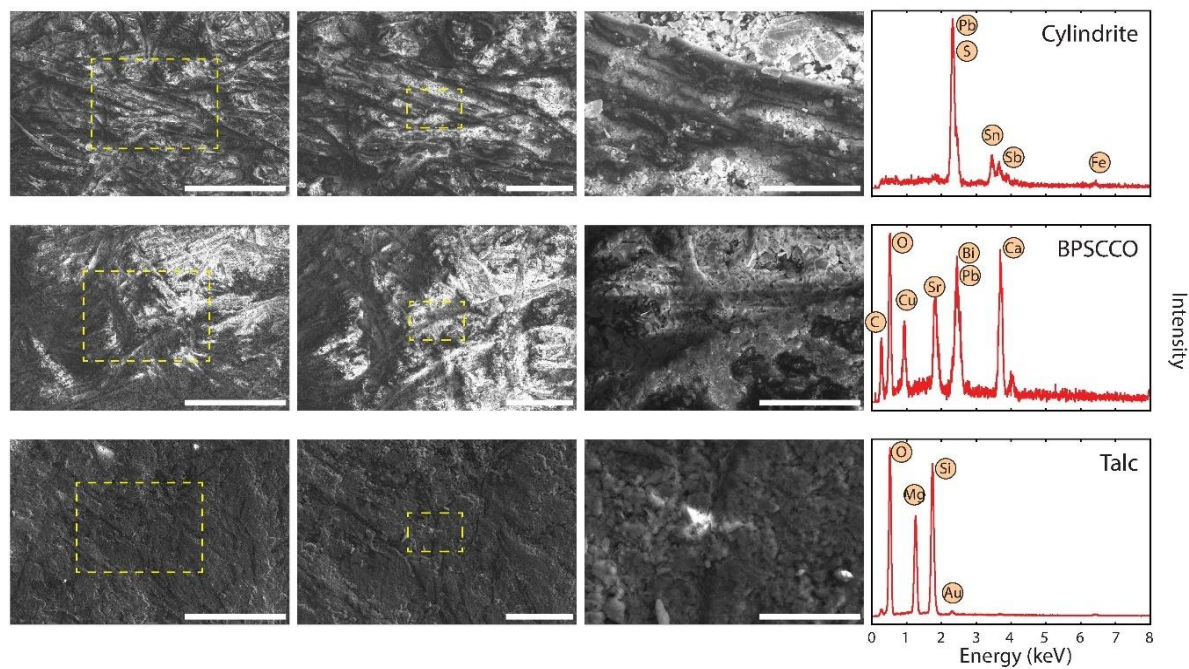

**Figure S8. SEM and EDX characterization of different films of vdW materials on paper.** Each row included 3 SEM images at increasing magnification and an EDX spectra of a different vdW material on paper (labelled in the EDX spectra panel). A thin layer of sputtered gold has been used in some highly resistive samples to avoid electrostatic charging effects; this might show up in the EDX spectra as a 'Au' peak. Some EDX spectra also show peaks associated to calcium presence. These peaks are due to the underlying paper and arise from the calcium carbonate (a bright white mineral with chemical formula $CaCO_3$) added as a filler in the paper pulp. Scale bars are 300 μm, 100 μm and 30 μm from leftmost panel to the rightmost panel.

**Raman characterization of vdW materials on paper:**

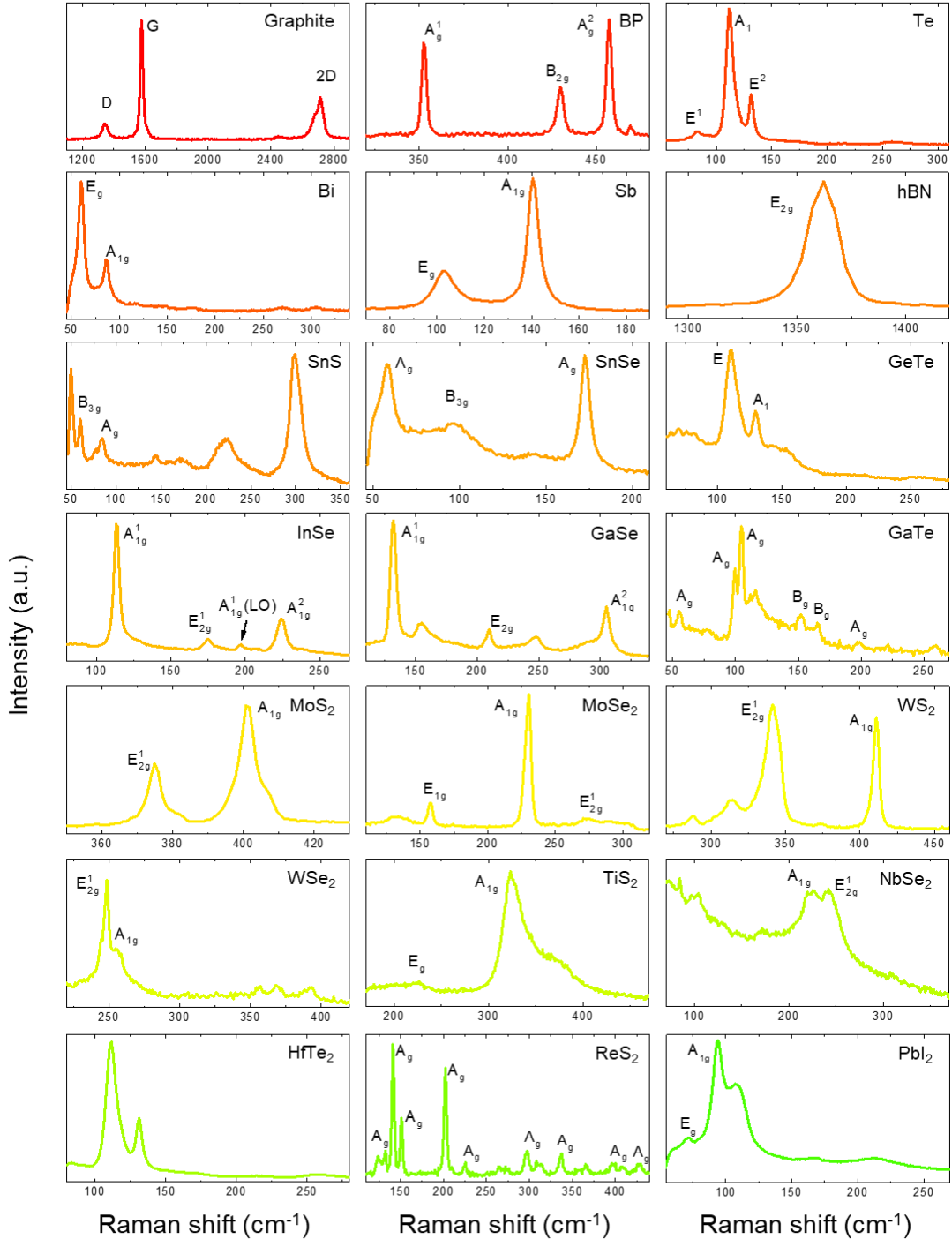

**Figure S9. Raman spectra measured on the different vdW materials on paper shown in Figure 2 of the main text. Bottom right: Raman spectra on bare paper.** The measured spectra are in excellent agreement to those reported in the literature for: graphite,[1] BP,[2] Te,[3,4] Bi,[5] Sb,[6,7] hBN,[8] SnS,[9–12] SnSe,[9,13] GeTe,[14] InSe,[15,16] GaSe,[17] GaTe,[18,19] MoS$_2$,[20,21] MoSe$_2$[22,23] WS$_2$,[24,25] WSe$_2$,[24–26] TiS$_2$[27,28] NbSe$_2$,[29] ReS$_2$[30,31] and PbI$_2$,[32,33] Note that the Raman spectrum of HfTe$_2$ is compatible with laser-induced damage or severe atmospheric aging.[34]

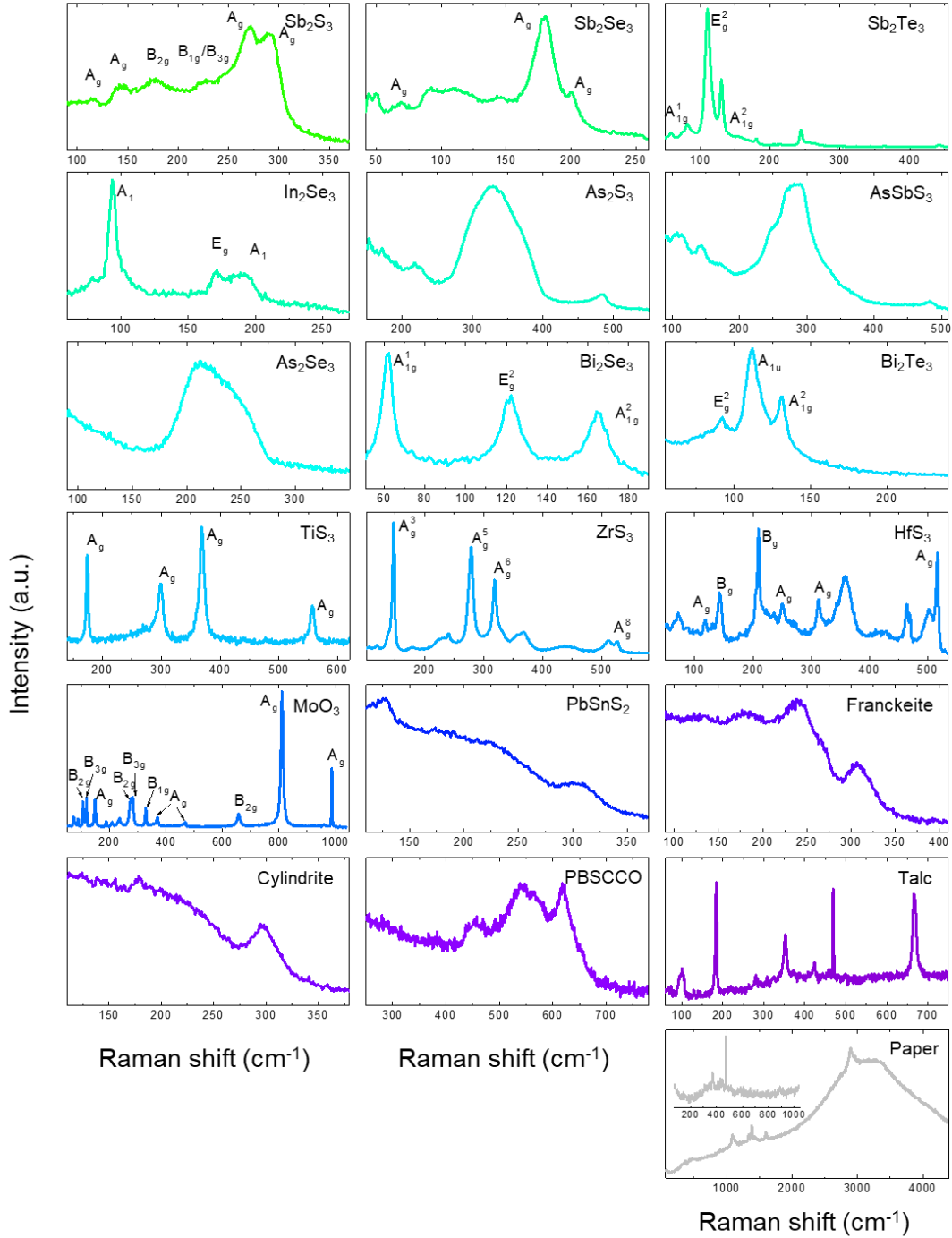

**Figure S10. Raman spectra measured on other vdW materials on paper.** The measured spectra are in excellent agreement to those reported in the literature for: $Sb_2S_3$,[35,36] $Sb_2Te_3$,[37,38] $Sb_2Te_3$,[39,40] $In_2Se_3$,[41,42] $As_2S_3$,[43–45] $AsSbS_3$,[46,47] $As_2Se_3$,[45,48,49] $Bi_2Se_3$,[50,51] $Bi_2Te_3$,[39,52] $TiS_3$,[53,54] $ZrS_3$,[55] $HfS_3$,[56] $MoO_3$,[57,58] franckeite,[59–61] cylindrite,[47] BPSCCO $(Bi_{1.6}Pb_{0.4}Sr_{1.6}Ca_{2.0}Cu_{2.8}O_{9.2+x})$[62,63] and talc[64]. The PbSnS$_2$ spectrum differs from previously published spectra. We attribute this to the fact that our PbSnS$_2$ source is natural teallite mineral and the PbSnS$_2$ spectra has a strong dependence on its exact composition and chemical impurities presence. [65,66] The lower right panel includes the Raman characterization of the standard copier paper substrate that presents a broad fluorescence.

**Transmittance spectra of the vdW materials not shown in the main text:**

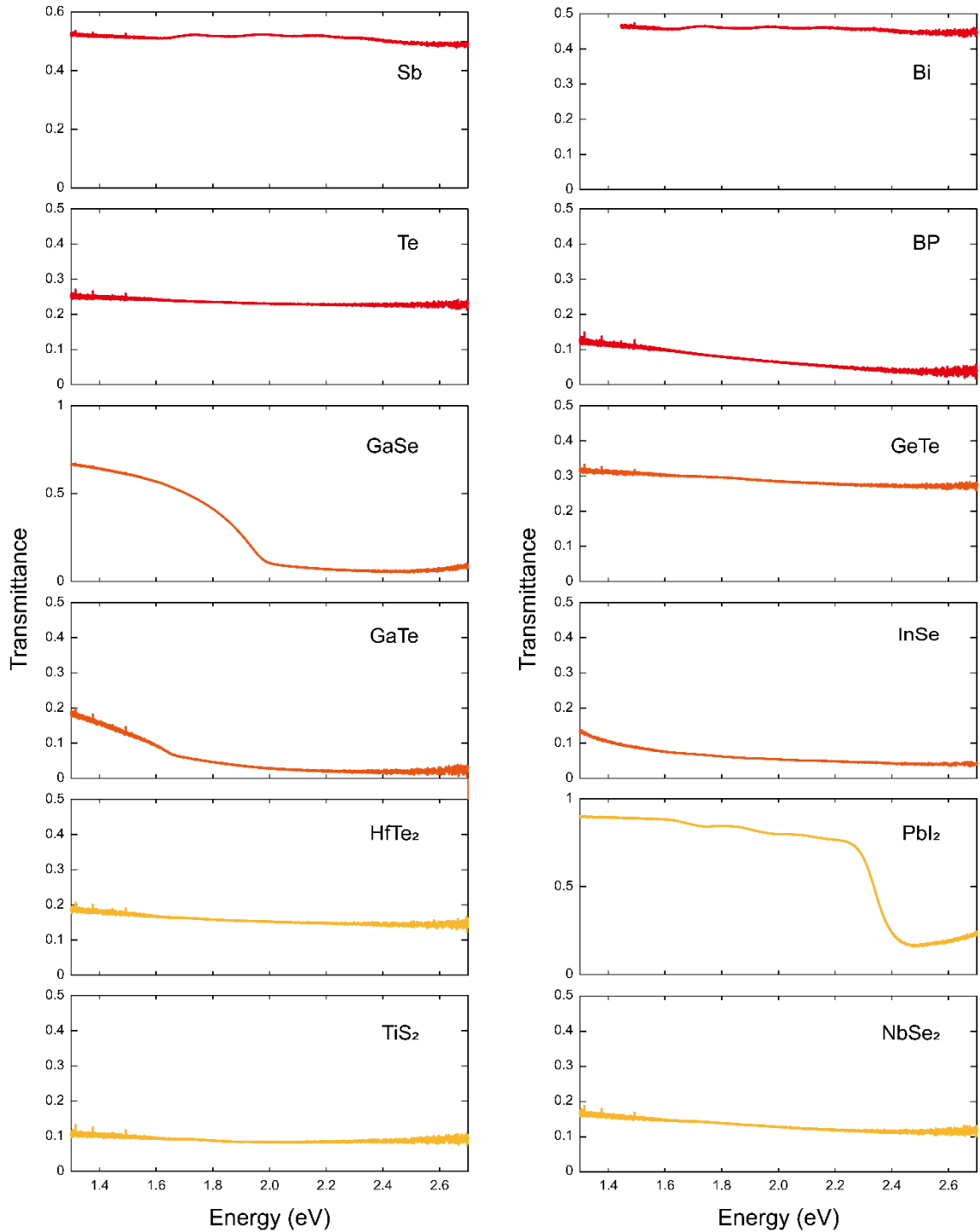

**Figure S11.** Transmittance spectra measured on different vdW materials on paper.

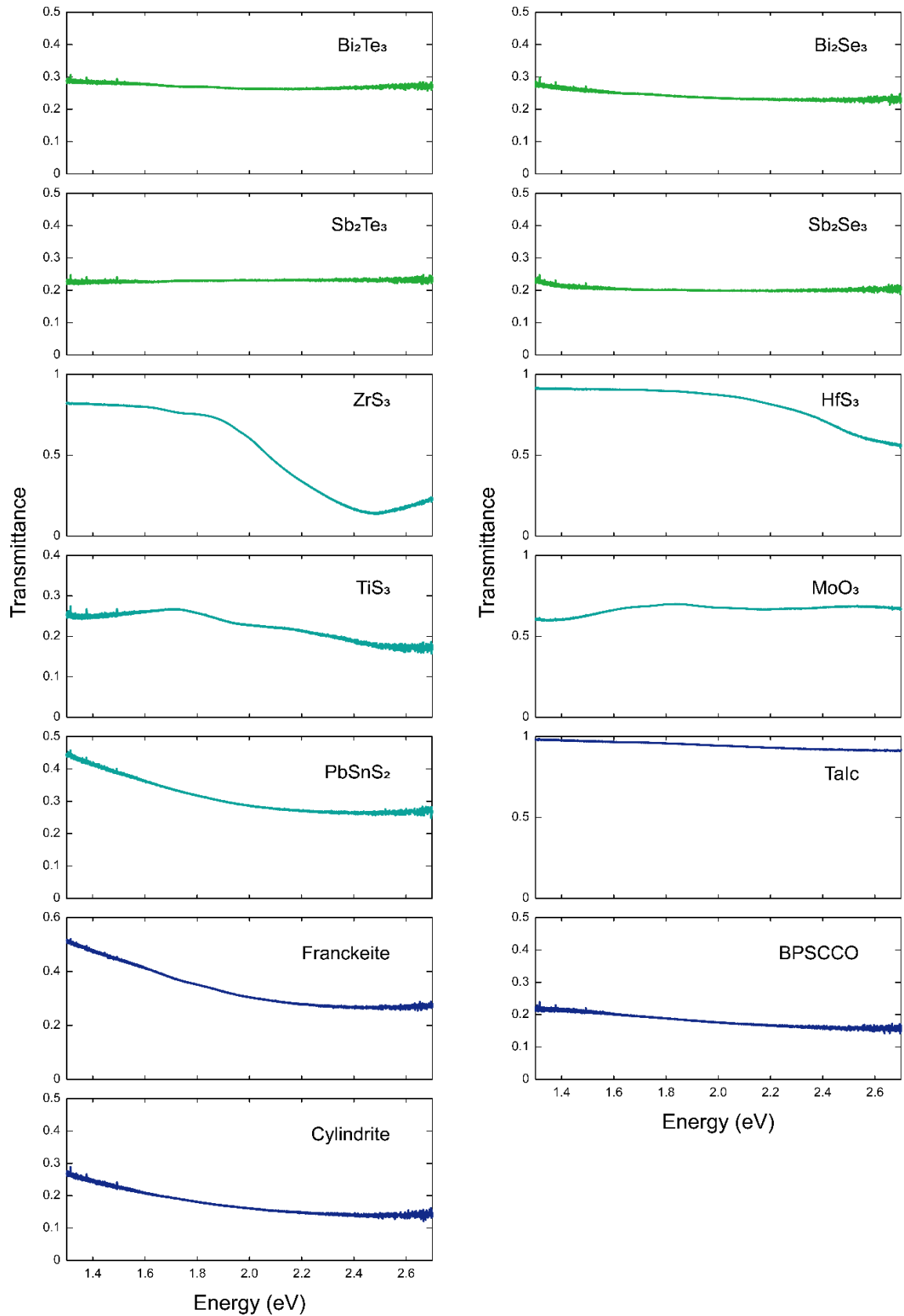

**Figure S12. Transmittance spectra measured on different vdW materials on paper.**

**Current *vs*. voltage characteristics for high voltage biasing:**

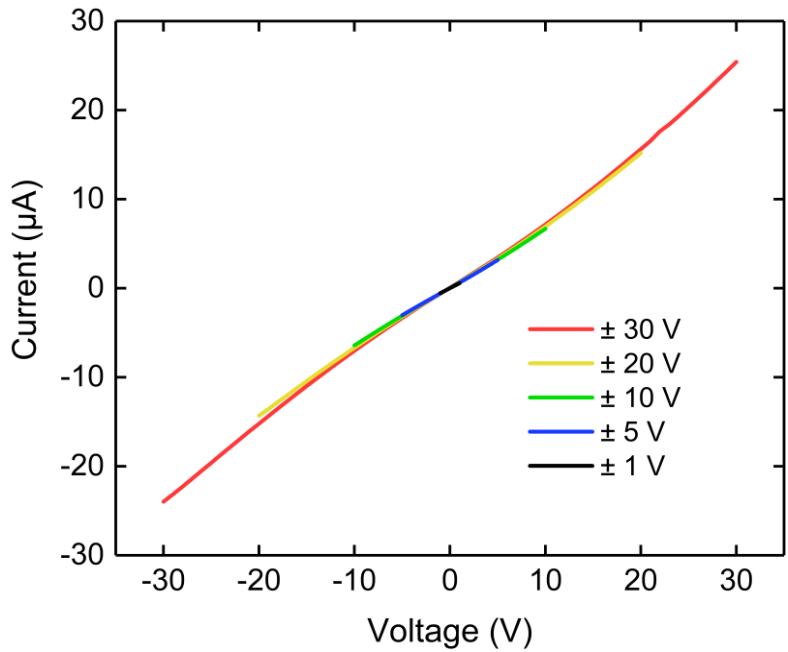

**Figure S13. Current vs. voltage characteristics for high voltage biasing.** *IV*s measured on a WS$_2$ device ($L$ = 2 mm, $W$ = 10 mm, graphite electrodes) at different maximum biases. We attribute the non-linearity observed for increasingly high bias window to self-heating effect.

**Electrical characterization of 14 vdW materials on paper:**

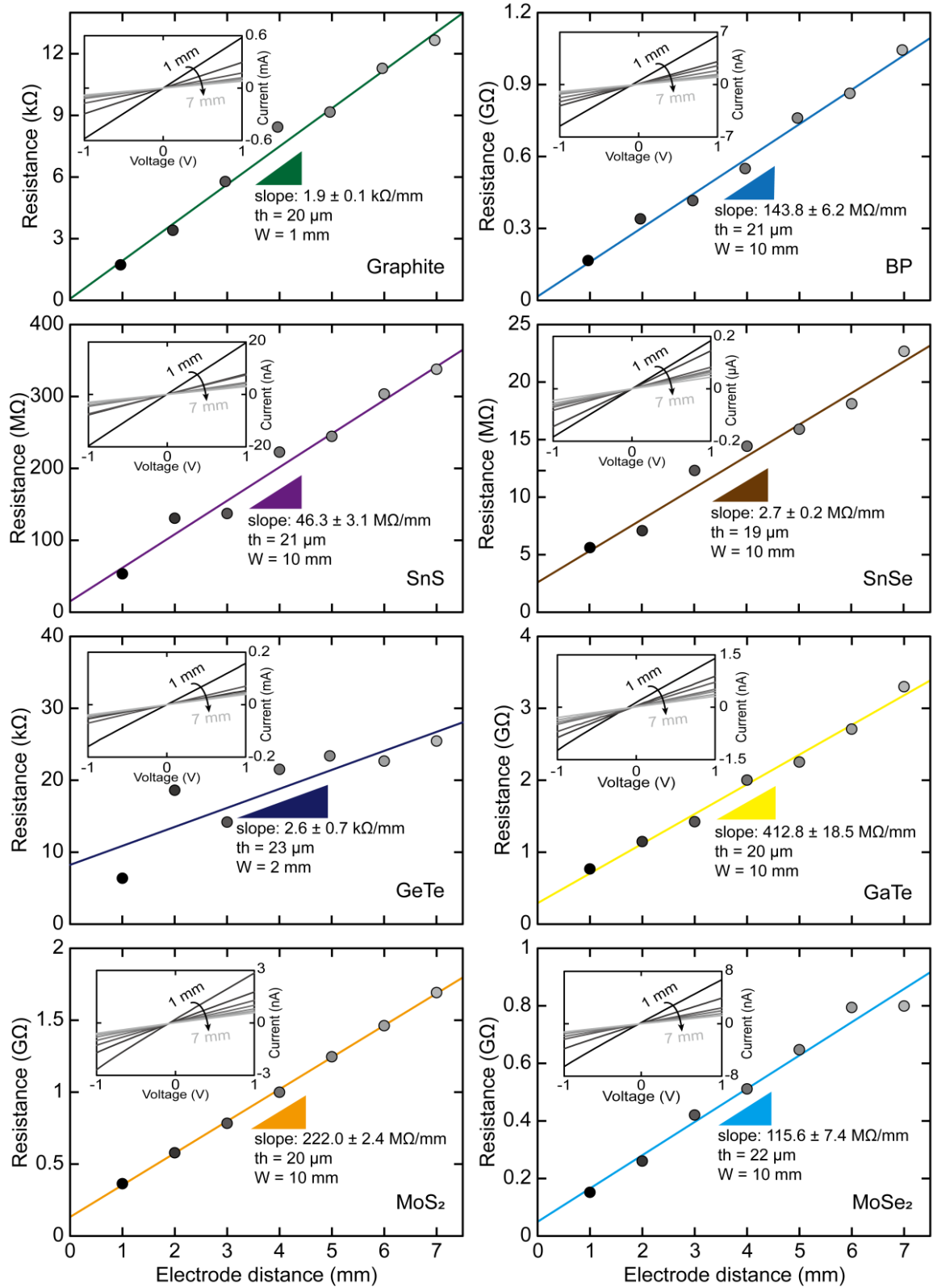

**Figure S14. Transfer length electrical transport measurements on devices based on different vdW materials.**

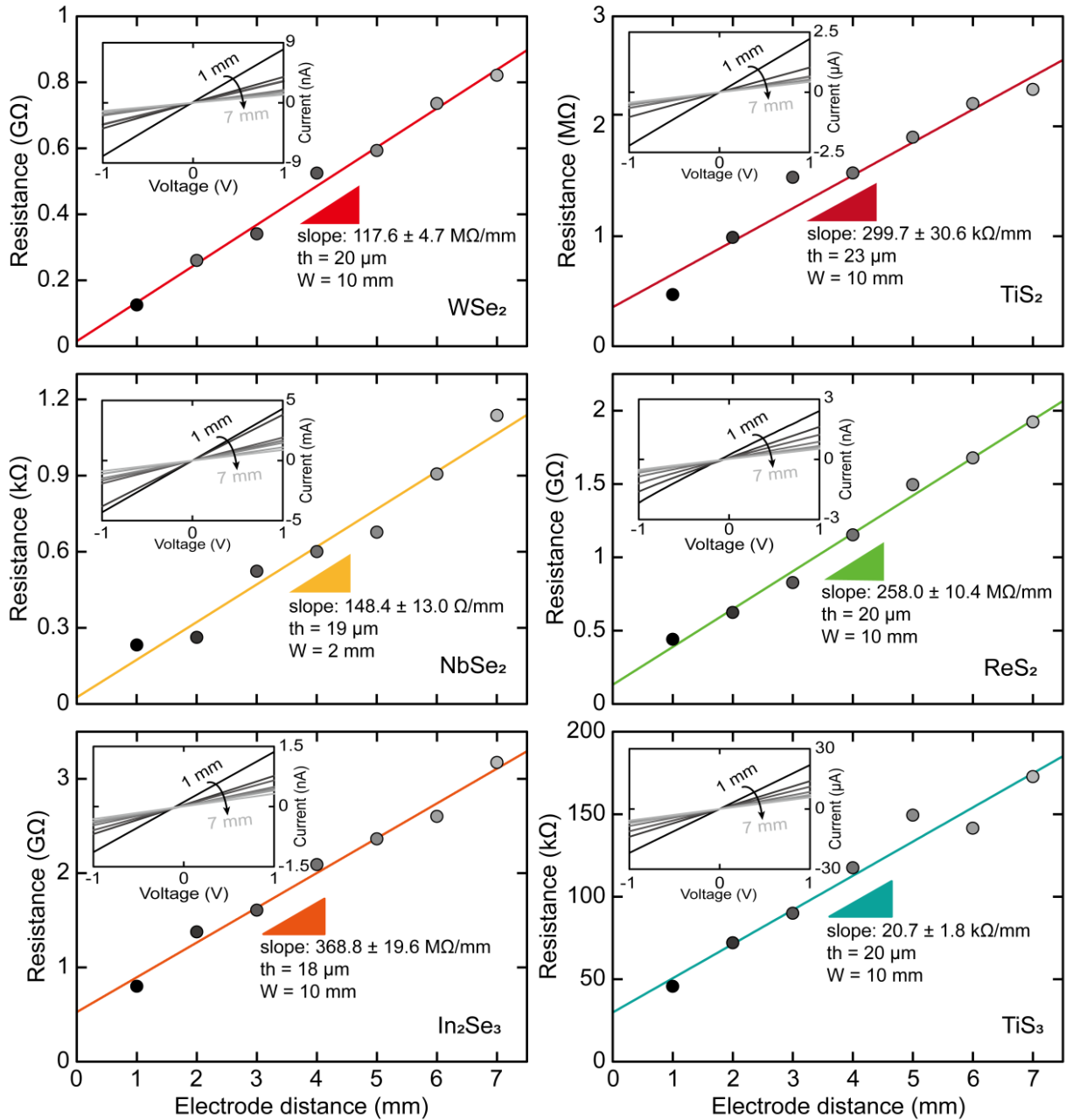

**Figure S15.** Transfer length electrical transport measurements on devices based on different vdW materials.

**Comparison of WS₂ devices from different material sources**

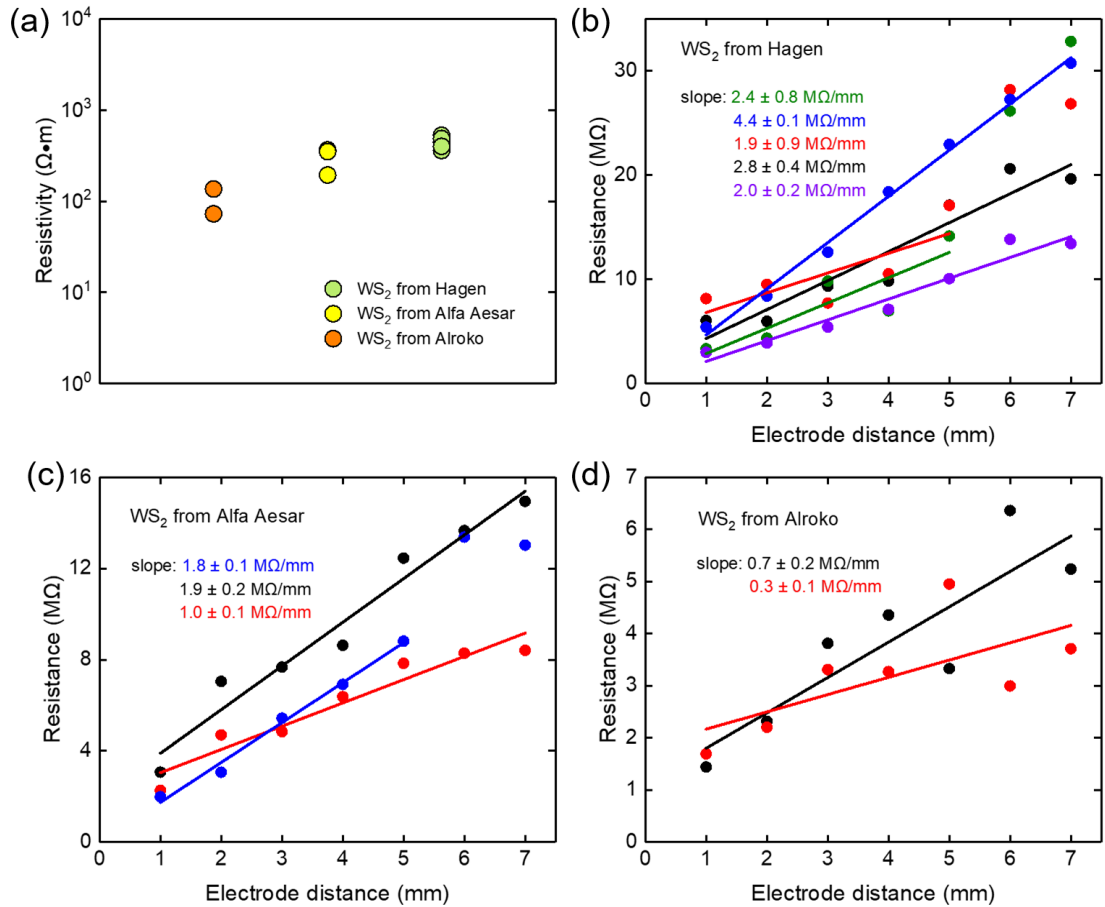

**Figure S16.** (a) Comparison of the resistivity values measured from WS₂ transfer length devices fabricated with WS₂ from different sources. (b) to (d) Transfer length electrical transport measurements on devices based on WS₂ from Hagen, Alfa Aesar and Alroko.

**Comparison of MoS₂ devices from different material sources**

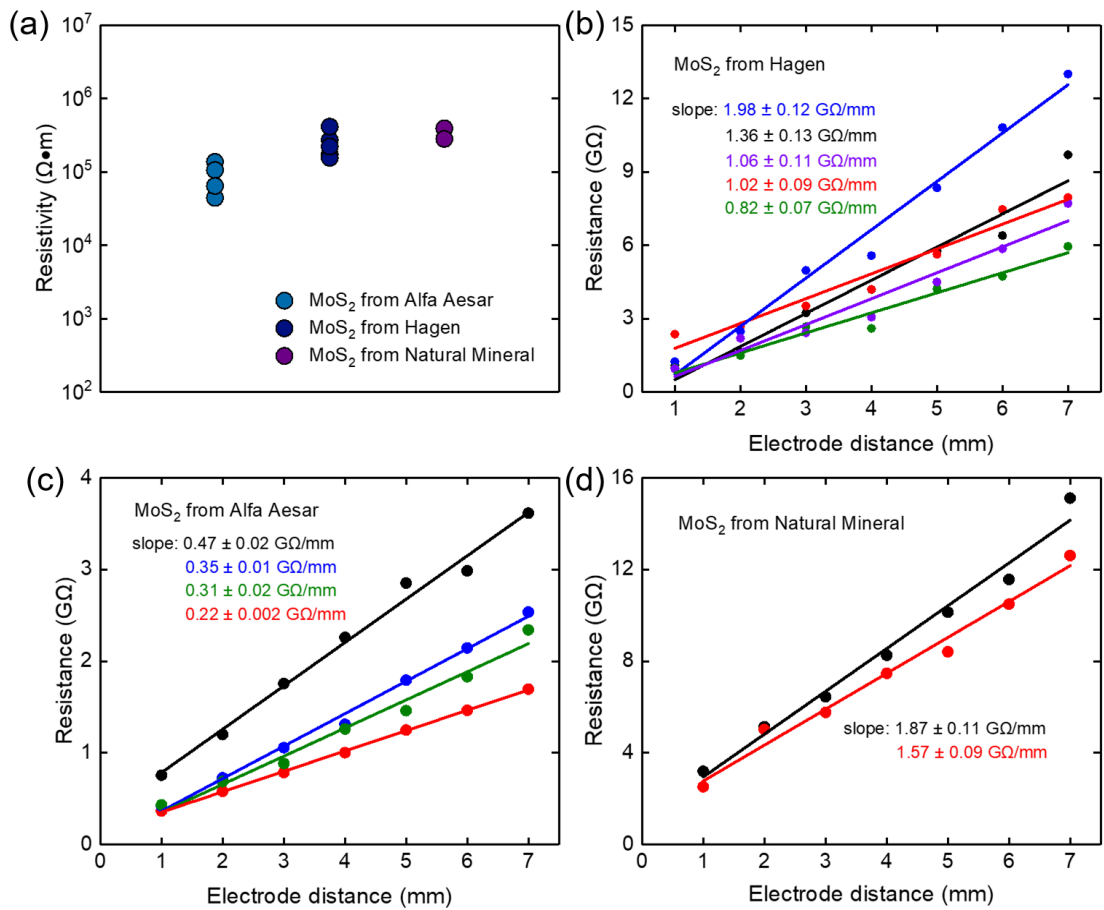

**Figure S17.** (a) Comparison of the resistivity values measured from MoS₂ transfer length devices fabricated with WS₂ from different sources. (b) to (d) Transfer length electrical transport measurements on devices based on MoS₂ from Alfa Aesar, Hagen and a natural molybdenite mineral.

**Comparison of graphite devices from different material sources**

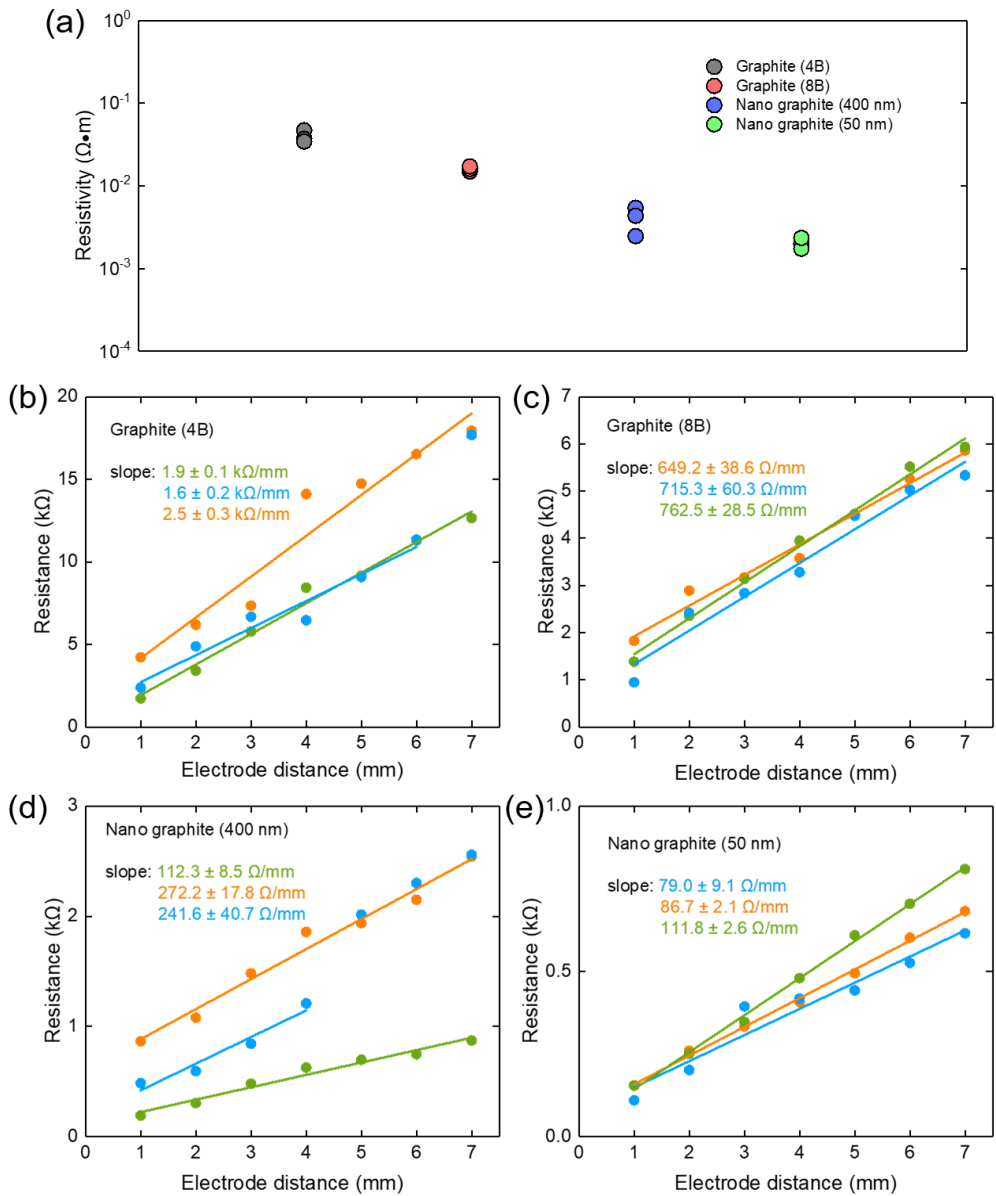

**Figure S18.** (a) Comparison of the resistivity values measured from graphite transfer length devices fabricated with graphite from different sources. (b) to (e) Transfer length electrical transport measurements on devices based on graphite from 4B Faber Castel pencil, 8B Faber Castel pencil, Lowerfricion Lubricants NanoGraphite 400 nm (MKN-CG-400) Lowerfricion Lubricants NanoGraphite 50 nm (MKN-CG-50).

**Simulation of a network of interconnected platelets through a random resistor network model:**

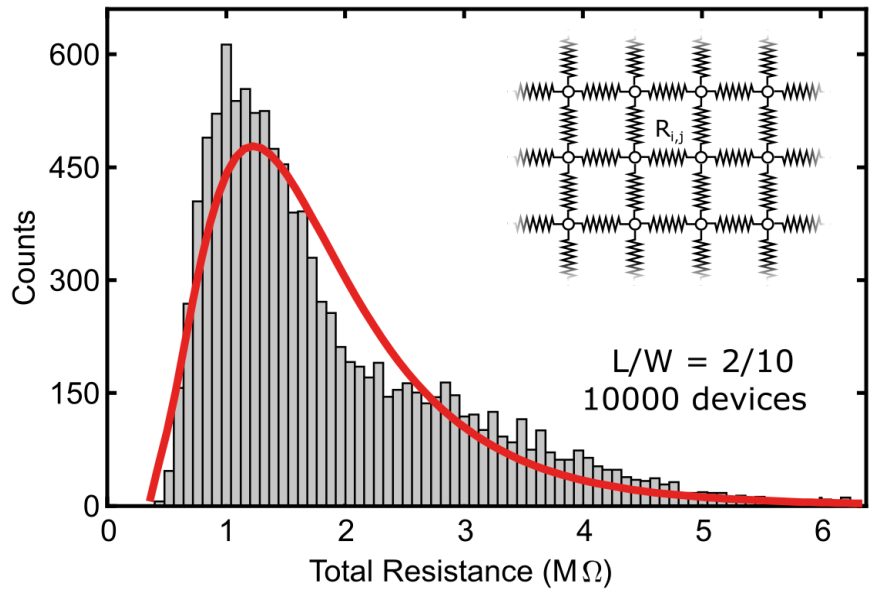

**Figure S19. Simulation of a network of interconnected platelets through a random resistor network model.** Histogram of the total resistance of 10000 different random resistors networks simulating 10000 channels of interconnected platelets with length/width aspect ratio equal to 2/10. The values of each resistance $R_{ij}$ in the network is randomly extracted from a binomial distribution with possible values $10^5$ Ω (simulating the intra-platelet resistance) and $10^{10}$ Ω (simulating the platelet-to-platelet hopping resistance). The red line is a fit to a lognormal distribution.

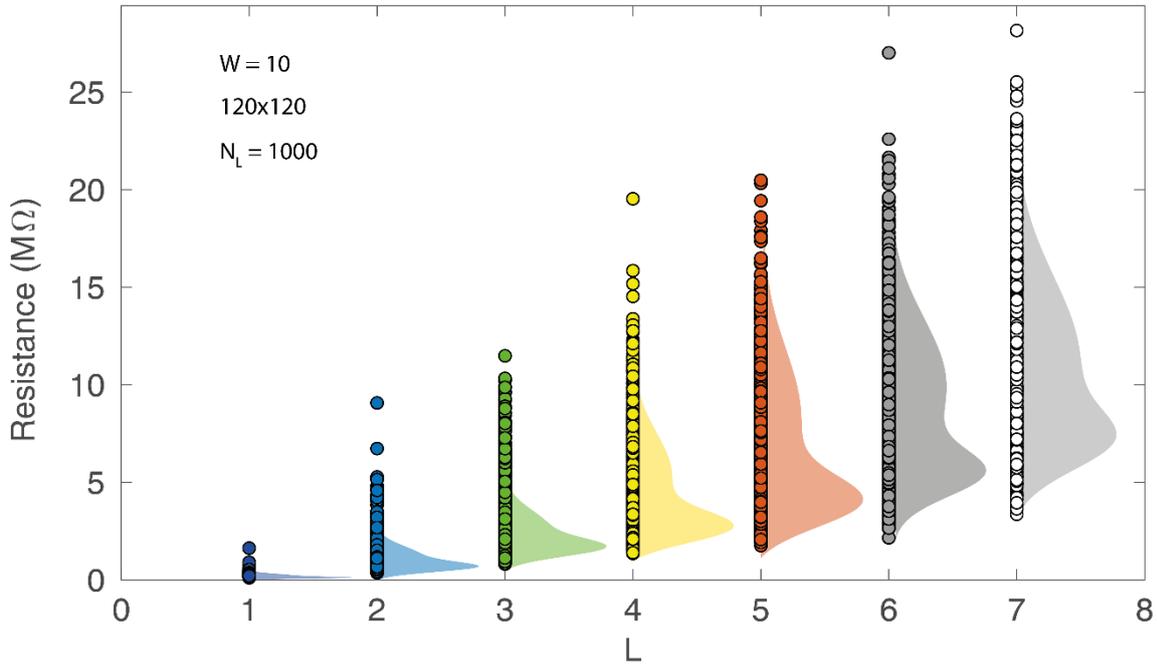

**Figure S20. Calculated resistance of a network of interconnected platelets through a random resistor network model considering different channel lengths.** For each channel length the resistance of 1000 different random network configurations are calculated. Besides the datapoints a histogram, built with the calculated resistance values for each channel length, is displayed. This illustrates how the resistance dispersion increases for longer channels and explains the observation of some transfer length characteristics measurements where there are outlier points.

**Histogram of the thickness of different vdW materials films on paper:**

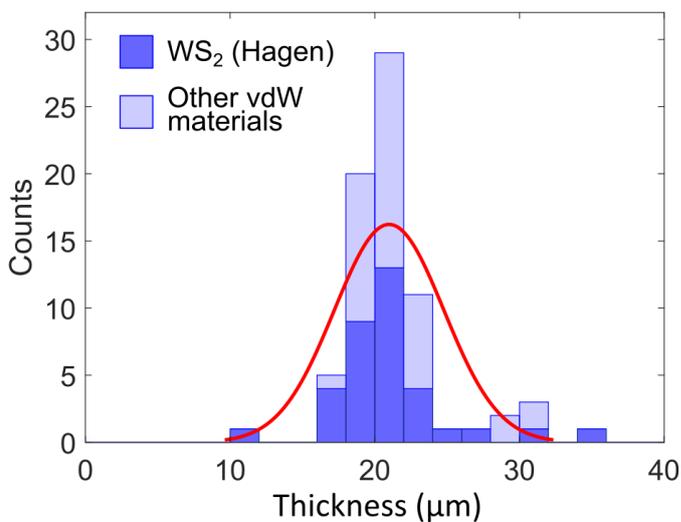

**Figure S21. Histogram of the thickness of different vdW materials films on paper. (dark blue bars)** Histogram built with the thickness of $WS_2$ Hagen films (×35). **(light blue bars)** Histogram built with the thickness of: $NbSe_2$ (×3), Graphite (×3), GeTe (×3), $TiS_2$ (×3), SnSe (×2), SnS (×1), BP (×1), $In_2Se_3$ (×1), $MoSe_2$ (×1), $WSe_2$ (×3), $ReS_2$ (×1), GaTe (×1), $MoS_2$ Alfa (×4), $MoS_2$ Hagen (×5), $MoS_2$ mineral (×2), $WS_2$ Alfa (×3) and $WS_2$ Alroko (×2) films.

## Supporting Information References: